\documentclass[a4paper,12pt]{article}
\usepackage{color}
\usepackage{verbatim}
\usepackage{braket}
\usepackage{amsthm}
\usepackage{amsmath}
\usepackage{graphicx}
\usepackage{amssymb}
\usepackage{esint}
\usepackage{subfigure}
\usepackage{color}
\usepackage{watermark}

\usepackage{anysize}
\marginsize{2cm}{2cm}{2cm}{2cm}

\makeatletter

\def\beq{\begin{equation}}
\def\eeq{\end{equation}}
\def\bea{\begin{eqnarray}}
\def\eea{\end{eqnarray}}
\def\nn{\nonumber}

\theoremstyle{plain}

  \theoremstyle{remark}

\begin{document}

\title{A Recipe for Constructing Frustration-Free Hamiltonians with Gauge and Matter Fields in One and Two Dimensions}
\author{Miguel Jorge Bernab\'{e} Ferreira$^{a}$\footnote{migueljb@if.usp.br}, ~Juan Pablo Ibieta Jimenez$^{a}$\footnote{pibieta@if.usp.br},\\ Pramod Padmanabhan$^{a}$\footnote{pramod23phys@gmail.com},~Paulo Te\^{o}tonio Sobrinho$^{a}$\footnote{teotonio@if.usp.br}} 

\maketitle

\begin{center}
{Departamento de F\'{i}sica Matem\'{a}tica Universidade de S\~{a}o Paulo - USP} \\
{\small
CEP 05508-090 Cidade Universit\'aria, S\~{a}o Paulo - Brasil
}

\end{center}

\begin{abstract}

State sum constructions, such as Kuperberg's algorithm, give partition functions of physical systems, like lattice gauge theories, in various dimensions by associating local tensors or weights, to different parts of a closed triangulated manifold. Here we extend this construction by including matter fields to build partition functions in both two and three space-time dimensions. The matter fields introduces new weights to the vertices and they correspond to Potts spin configurations described by an $\mathcal{A}$-module with an inner product. Performing this construction on a triangulated manifold with a boundary we obtain the transfer matrices which are decomposed into a product of local operators acting on vertices, links and plaquettes. The vertex and plaquette operators are similar to the ones appearing in the quantum double models (QDM) of Kitaev. The link operator couples the gauge and the matter fields, and it reduces to the usual interaction terms in known models such as $\mathbb{Z}_2$ gauge theory with matter fields. The transfer matrices lead to Hamiltonians that are frustration-free and are exactly solvable. According to the choice of the initial input, that of the gauge group and a matter module, we obtain interesting models which have a new kind of ground state degeneracy that depends on the number of equivalence classes in the matter module under gauge action. Some of the models have confined flux excitations in the bulk which become deconfined at the surface. These edge modes are protected by an energy gap provided by the link operator. These properties also appear in ``confined Walker-Wang'' models which are 3D models having interesting surface states. Apart from the gauge excitations there are also excitations in the matter sector which are immobile and can be thought of as defects like in the Ising model. We only consider bosonic matter fields in this paper.  

\end{abstract}

\section{Introduction}

Lattice systems with gauge and matter fields is a recurring theme in high energy and condensed matter physics. They are a useful way of regularizing field theories in particle physics where the standout example is lattice QCD, a widely studied field~\cite{Kreutz}. The gauge groups involved in these programs are compact, continuous non-Abelian groups like $SU(3)$ and $SU(2)$ and Abelian groups like $U(1)$. Whereas in the context of condensed matter physics such lattice systems are used to model magnetism in solids and crystals and study various phases of strongly correlated electrons most of which are typically low energy phenomena. One of the prominent examples is the Hubbard model which is an interacting microscopic model used to describe a variety of strongly correlated phenomena~\cite{Hubbard}. In these examples the gauge groups considered are finite discrete groups such as the Abelian $\mathbb{Z}_n$. The classic examples using such Abelian groups as degrees of freedom are the Ising chain, Heisenberg model and the Potts model~\cite{Spinsystems}. An important connection between lattice spin systems and gauge theories was given in the famous article by Kogut~\cite{Ko1, Ko2}. They study in detail the connection between a $d$-dimensional quantum spin system and the $d+1$-classical system via the transfer matrix method. This gives the correspondence between the two-dimensional quantum Ising model and three dimensional $\mathbb{Z}_2$ lattice gauge theory.  

The partition functions of such lattice systems can be found out in a systematic way by looking at them as state sum models. These state sum models are used in several areas of mathematics and physics, including statistical mechanics \cite{baxter}, random matrices \cite{random}, knot theory \cite{kauffman}, lattice gauge theory \cite{oeckl} and quantum gravity \cite{barrett}. Such state sum models are defined based on a combinatorial decomposition of a manifold such as a lattice or a triangulation, which can be interpreted to be the space-time in a physical picture. In these models the degrees of freedom live on the vertices and/or edges of the lattice. In a more general setting one could associate local states on even faces and volumes. We restrict our attention to those living only on the vertices and edges. A state in such a system is the tensor product of the configuration in each local edge and vertex. Local weights are associated to the vertices edges and faces of the triangulated lattice. These weights can be thought of as tensors with indices which can be raised and lowered. These tensors are then allowed to contract and thus one obtains a number corresponding to the partition function when this construction is carried out for a closed triangulated manifold. 

These constructions have been carried out in the past for finding 3-manifold invariants which were also interpreted as the partition functions of certain physical models. They were especially useful in constructing the partition functions of Topological Quantum Field Theories (TQFTs). An important example is the 3D Dikgraaf-Witten invariant~\cite{DikWit} which furnished all topological lattice gauge theories. Hamiltonian realizations of these invariants in two dimensions is given in~\cite{YSWu}. Another important example is the Turaev-Viro type of TQFTs in 3D~\cite{Tu}. These are realized by the Levin-Wen models or the string-net Hamiltonians~\cite{stringLevin} in two dimensions. These are exactly solvable models made up of commuting projectors describing degrees of freedom belonging to a unitary fusion category located on the edges of the lattice. They realize anyons as low energy excitations and provide a general class of long-ranged entangled topologically ordered phases in two dimensions. The Turaev-Viro invariant are known to be equivalent to the Chain-Mail link invariant~\cite{cmLink}. It was shown in~\cite{burnell1} how one can obtain the string-net models from the Chain-Mail link invariant which is also a knot invariant. 

In the spirit of such computations we showed in~\cite{pp1} that the partition function of Quantum Double Hamiltonians (QDM) of Kitaev~\cite{Kit, Agu} can be obtained by a deformation of Kuperberg's 3-manifold invariants~\cite{Kuper1}. The QDM Hamiltonians realize lattice gauge theories based on an involutory Hopf algebra $\mathcal{A}$. Lattice gauge theories based on a group $G$ is a special case of this as $\mathcal{A}=\mathbb{C}(G)$ is an involutory Hopf algebra. When $G=\mathbb{Z}_2$ we obtain the toric code model. Thus the quantum double Hamiltonian realizes a representation of the quantum double of these involutory Hopf algebras. This is a general recipe to construct quasi triangular Hopf algebras from a given Hopf algebra. These quasi triangular structures are governed by a $R$-matrix which form a representation of the braid group in two dimensions and hence help us find anyonic solutions~\cite{Majid}. In~\cite{pp1} we embed such quantum double models in a bigger parameter space namely the full parameter space of lattice gauge theories in three dimensions. For particular choices of the parameters we obtain the solvable models which describe the same phase as the QDM phase of Kitaev. For more general parameters we go away from this phase in a manner similar to the effect of perturbations which can be thought of as magnetic fields in the simple cases of the group algebra. By studying it in a bigger parameter space new topological and quasi-topological phases were also shown to exist~\cite{pp1} for these lattice gauge theories. These phases were understood by analyzing the Hamiltonians derived from the transfer matrices of these lattice gauge theories. These were constructed by carrying out Kuperberg's prescription on a triangulated manifold with boundary. These transfer matrices are a product of local operators acting on the vertices, plaquettes and links. The vertex and plaquette operators are precisely the ones appearing in the QDM Hamiltonian~\cite{Kit, Agu}. The ones acting on the links are like the perturbations to the QDM Hamiltonian. The trace of these transfer matrices coincide with the Kuperberg 3-manifold invariant. In this regard~\cite{pp1} is a significant extension of Kuperberg's constructions as it is not limited to just obtaining 3-manifold invariants. By also parametrizing the transfer matrix we go away from the topological limits and recover the topological invariants for special parameters which incidentally also give the topologically ordered models.

The QDM and the string-net models are all examples of Hamiltonian realizations of lattice gauge theories based on involutory Hopf algebras and quantum groupoids or weak Hopf algebras~\cite{LChang} respectively. The gauge fields in these examples live on the edges of the three dimensional lattice. These are examples of long-ranged entangled (LRE) phases which form a large chunk of known topologically ordered phases~\cite{NayakReview}. The ``other'' chunk of known topological phases are the short-ranged entangled ones. These include the Symmetry Protected Topological (SPT) phases which are interacting bosonic phases which have edge states protected by a global symmetry group~\cite{wenSPT}. Conventional topological insulators~\cite{BernevigBook} are non-interacting fermionic versions of these phases. Such phases are known to exist in all dimensions, however the physically interesting ones are the ones in one, two and three dimensions. Exactly solvable effective models describing such phases have been studied in~\cite{ls1,ls2}. Gauging the global symmetries of these systems lead to fractional excitations in the bulk. These help mimic the behavior of fractional quantum Hall (FQH) states~\cite{fqh}.

Apart from these two broad classifications of topological phases, the LRE phases were further subdivided when they are decorated with an additional symmetry giving rise to symmetry enriched topological (SET) phases~\cite{wang, hughes1, hughes2}. These phases were studied in the mathematical framework of $G$-crossed braided fusion categories~\cite{galindo}. Exactly solvable lattice models for realizing SET phases with global symmetries were done in~\cite{herm1, herm2, fid}. Gauging these global symmetries leads to new topologically ordered phases starting from the parent topological phase. Such phases were obtained locally in exactly solved Hamiltonians~\cite{bombin, wen2, ppm}. They were realized in more realistic systems, such as bilayer FQH states, by introducing dislocations and the accompanying branch cuts~\cite{maisham}. 

A common feature of the above systems is the existence of a global symmetry. In the case of the SPT phases these global symmetries are realized by using matter fields living on the vertices with no link degrees of freedom~\cite{wenSPT} and for the SET phases the global symmetry is realized by placing the system on an enlarged lattice~\cite{herm1, herm2, fid} or by introducing dislocations in the lattice~\cite{maisham}. Thus it is desirable to go beyond the pure gauge systems in~\cite{pp1} by implementing an additional global symmetry which may help one obtain the SET and SPT phases in a systematic manner. We do this by extending the construction in~\cite{pp1} by including matter fields on the vertices and construct the corresponding transfer matrices of systems with gauge and matter fields. By using this transfer matrix we obtain exactly solvable models of such systems which also includes the interaction between the gauge and matter fields. These Hamiltonians are frustration free~\cite{spiros} and possess both a global symmetry, in the matter sector and a local symmetry in the gauge sector. They are frustration free in the sense that the Hamiltonian is a sum of terms such that the ground states of the full Hamiltonian are the lowest energy states of each individual term. The global ground states are also local ground states and hence there is no frustration or energy increase when all the terms in the Hamiltonian are included. A Hamiltonian which is a sum of commuting projectors satisfies these conditions and we shall see that the Hamiltonians constructed in this approach are of this form.

Before we provide a description of the construction of these transfer matrices we briefly write down some properties of the systems obtained from these transfer matrices. As noted before the Hamiltonians are exactly solvable and the spectrum is gapped. The main input in building these systems is the gauge group and the matter module which is acted upon by the gauge group and the choice of the representation of the gauge group on the matter module. Depending on this choice we have can obtain a variety of systems. A common feature to all of them is the ground state degeneracy which is no longer a topological invariant like in the pure gauge case~\cite{pp1} but rather develops another kind of degeneracy which depends on the number of equivalence classes of the matter module under the gauge action. Depending on the choice of the action of the gauge group the system can also develop a topological degeneracy in addition to the one coming from the different equivalence classes. These systems have excitations in both the gauge and the matter sectors. Unlike the deconfined excitations of the LRE phases, only the charge excitations are deconfined in this model. Depending on the choice of the action of the gauge fields on the matter degrees of freedom the system may or may not have deconfined fluxes. When the fluxes are confined, they are done so by string tension terms provided by operators acting on links as we shall see soon. However when these systems are placed on a manifold with a boundary the fluxes get deconfined and become edge modes which are protected by an energy gap provided by the link operators. This feature will be explained with a specific example in the text. This property of confined bulk excitations and deconfined excitations on the edges has been observed in confined Walker-Wang models~\cite{ww} elaborated in~\cite{ash1, ash2, ash3}. The confined Walker-Wang models are systems in 3D, which when placed on manifolds with boundaries result in topologically ordered surface states. The examples in this paper achieve this in two dimensions through gapped edge modes.

We briefly describe our method to construct the transfer matrix of a lattice system with gauge and matter fields in two and three dimensions thus corresponding to quantum lattice models in one and two dimensions respectively. For convenience we only work with the case of finite groups though our methods can be extended to the case of continuous compact groups and in general to involutory Hopf algebras. Thus our focus is only on models which are pertinent for condensed matter systems. In particular we will be interested in finding exactly solvable Hamiltonians which act as effective descriptions of interacting systems of gauge and matter fields in one and two dimensions. We will explain the procedure for the three dimensional case. The procedure can be easily followed in the two dimensional case as well and is included as an appendix to this paper.

The construction proceeds as follows. The data needed to define a partition function is a triangulated, closed 3-manifold which we take to be of the form $\Sigma\times S^1$, where $\Sigma$ is the two dimensional spatial slice, an involutory Hopf algebra $\mathcal{A}$ which make up the gauge degrees of freedom located on the edges of the triangulated 3-manifold and a $n$-dimensional vectorial space $H_n$, which carries a representation of $\mathcal{A}$\footnote{In other words, $H_n$ is a left $\mathcal{A}$-module.} and has a co-structure and an inner product. The elements of $H_n$ make up the matter fields which are located on the vertices of the triangulated 3-manifold. The invariant is built by contracting local weights associated to vertices, edges and faces. These local weights are built out of the structure constants of the involutory Hopf algebra $\mathcal{A}$ and the co-algebra $H_n$. These structure constants from $\mathcal{A}$ are given by tensors $m_{ab}^c$, $\Delta_c^{ab}$ and $S_a^b$. They correspond respectively to the multiplication (product) map, coproduct map and the antipode. The structure constants from $H_n$ are given by tensors $t_\gamma^{\alpha\beta}$ and $\mu_a^{\alpha\beta}$ where the former is similar to the coproduct for coalgebras and it arises from the costructure of $H_n$ and the latter is the map which says how the gauge fields act on the matter fields, that is it is the map which furnishes a representation of $\mathcal{A}$ on $H_n$.  In the structure constants listed the Latin alphabets index the gauge degrees of freedom and the Greek alphabets index the matter degrees of freedom. The scalar constructed out of these structure constants can be thought of as the result of contracting a three dimensional tensor network where we now consider the local weights as local tensors associated to the various parts of the three dimensional lattice. The partition function $Z$ is parameterized by $z$ and $z^*$, elements of the centers of the algebra $\mathcal{A}$ and its dual, $m_V$ an element of $H_n$ and $G$ the inner product in $H_n$. Thus we obtain $Z\left(\mathcal{A}, H_n, z, z^*, m_V, G\right)$ which is in general a scalar and need not be topologically invariant.  

Obtaining an operator from this scalar is a natural step when we carry out the above procedure on a 3-manifold with boundary. This operator is precisely the transfer matrix as it's trace is the partition function. The 3-manifold we consider is $\Sigma\times [0, 1]$ where as before $\Sigma$ is the two dimensional spatial slice and the unit interval is along the third direction which we can think of as the Euclidean time direction. Thinking of the partition function as a contracted tensor network we now have non-contracted indices on the spatial slices $\Sigma\times \{0\}$ and $\Sigma\times \{1\}$. This results in the transfer matrix $U$ for the lattice system. However since we distinguish spatial and ``time'' directions in this construction we have more parameters for $U$. This means that the weights associated to the two directions are not necessarily the same. Denoting the space and time directions by $S$ and $T$ respectively we now have the fully parametrized transfer matrix as $U\left(\mathcal{A}, H_n, z_S, z_T, z_S^*, z_T^*, m_V, G_S, G_T\right)$. We show how this global operator can be written as a product of local operators acting on vertices, links and plaquettes. We obtain
\begin{eqnarray} U\left(z_S, z_T, z_S^*, z_T^*, m_V, G_S, G_T\right) & = & \prod_p B_p(z_S) \prod_l \tilde{C}_l(G_S) \prod_l\left(T_l(z_S^*)D_l L_l(z_T) \right) \nn \\
 & & \prod_v \tilde{V}_v(G_T)Q_v(m_v)\prod_v A_v(z_T^*) \end{eqnarray} 
where $v$, $l$ and $p$ denote the vertices, links and plaquettes respectively. 

This operator is very general and encompasses a wide variety of interacting lattice models with gauge and matter fields. However for specific values of the parameters, namely
\begin{eqnarray}
z_T &= & \eta \;,\nn \\
z_S^* &= & ^\mathcal{A}\epsilon \;,\nn \\
m_V &= & ^{H_n}\epsilon \;,\nn \\
(G_T)_{\alpha\beta} &= & \delta(\alpha,\beta) \;,\nn
\end{eqnarray}
where $\eta$ and $^\mathcal{A}\epsilon$ are the unit and co-unit of the algebra $\mathcal{A}$ and $^{H_n}\epsilon$ is the counit of $H_n$, we obtain exactly solvable models. The set of parameters we will work with in this paper are $z_S$, $z_T^*$ and $G_S$ which give us 
\beq U(z_S, z_T^*, G_S) = \prod_p B_p(z_S) \prod_l \tilde{C}_l(G_S) \prod_v A_v(z_T^*).\eeq
These give us models which are similar to the toric code in the sense of a frustration-free Hamiltonian made up of commuting projectors and having long ranged entangled ground states, but including matter fields. This Hamiltonian is given by
\beq H = -\sum_v A_v -\sum_p B_p -\sum_l C_l \eeq
where the vertex operators are gauge transformations just as in the QDM case, the plaquette operators measure gauge fluxes around a plaquette, and the link operator which is a gauged term with Potts-like nearest neighbor interactions. In other words, $C_l$ couples the gauge configurations at the links with the Potts spins located at the vertices.

The contents of this paper are organized as follows. In section 2 we describe the construction of the transfer matrix in $2+1$ dimensions in a systematic manner starting from the method to construct the weights/local tensors using the structure constants of the input algebra for the gauge algebra and its corresponding vector space carrying it's representation. The partition function resulting from this assignment is briefly sketched before the transfer matrix is obtained from this using the splitting procedure. The splitting procedure leads to the transfer matrix written as a product of local operators. All these local operators are obtained at the end of section 3 as tensor networks, written down in the Kuperberg notation~\cite{Kuper1}. The algebraic expressions for these operators are given for the gauge algebra being the group algebra and the vector spaces being the ones carrying their representations, in section 4. The input algebra can be more general than group algebras with group algebras being most relevant for condensed matter physics. Some examples of exactly solvable models obtained from this construction are presented in section 5. A brief outlook makes up section 6. There are two appendices intended as a supplement to the main text. Appendix A furnishes the details of the basic input for our construction. In appendix B we show the same construction in $1+1$ dimensions which produces quantum models in $1D$. These can be used to obtain spin chain models and hence completes our formalism for physically relevant instances. Finally in appendix C we look at two examples of exactly solvable models in one dimension and find them to have interesting properties which warrant further study.

\section{Constructing the Transfer Matrix}

\begin{figure}[h!]
\centering
\subfigure[A piece of a $(2+1)$D square lattice.]{
\includegraphics[scale=1]{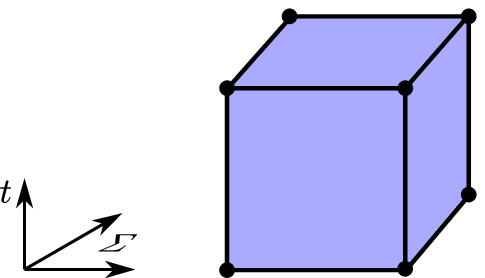} \label{lattice1-a}
}
\hspace{2.5cm}
\subfigure[Small pieces of which the $(2+1)$ lattice is made of.]{
\includegraphics[scale=1]{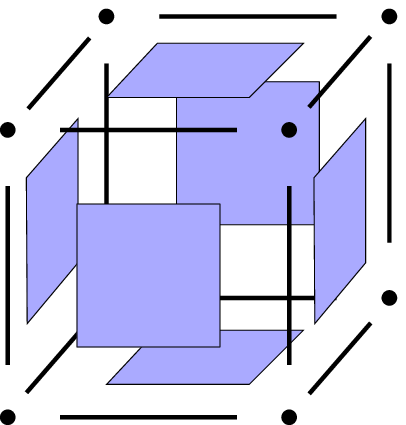} \label{lattice1-b}
}
\caption{Square lattice and its pieces.}
\label{lattice1}
\end{figure}

In order to define our model we define a lattice $\mathcal{L}$ which is a discretization of some $(2+1)-$manifold $\Sigma\times S^1$, where $\Sigma$ is some compact $2-$manifold such that $\partial \Sigma = \emptyset$. For the purpose of this work it is enough to consider $\mathcal{L}$ a square lattice as the one shown in figure \ref{lattice1-a}. This lattice is constructed by gluing vertices, links and faces as shown in figure \ref{lattice1-b}.
The model we build has degrees of freedom associated to gauge fields (living on the links) and degrees of freedom associated to matter (living on the vertices). They are quite general in the sense that it can support models such as the quantum double models which includes the toric code, besides models in other phases of matter. The Hamiltonian operator is the one which  propagates the states along the time direction and it is made up of a set of projectors operators which act on specific parts of the lattice. In this paper we will show how to get the Hamiltonian operator in the language of tensor networks. The way we proceed is very similar to the way we have done so in \cite{pp1}. We start with a partition function $Z$ written in the formalism of the state sum model which leads to a one step evolution operator $U$ such that \footnote{For brevity in the partition function expression we omit the dependence on the lattice ($\mathcal{L}$), on the algebra ($\mathcal{A}$) and on the $H_n$ space. So by $Z\left(z_\textrm{S},z_\textrm{T},z_\textrm{S}^*,z_\textrm{T}^*,m_V,G_S,G_T \right)$ we actually mean $Z\left(\mathcal{L},\mathcal{A},H_n,z_\textrm{S},z_\textrm{T},z_\textrm{S}^*,z_\textrm{T}^*,m_V,G_S,G_T \right)$.}
$$Z\left(z_\textrm{S},z_\textrm{T},z_\textrm{S}^*,z_\textrm{T}^*,m_V,G_S,G_T \right)=\textrm{tr}\left( U^N \right)\;,$$
where $z_\textrm{S},z_\textrm{T},z_\textrm{S}^*,z_\textrm{T}^*$ are elements of the center of algebra and co-algebra which play role of parameters of the model, the subscript S and T means spacelike and timelike parameters while $N$ is the number of steps in the time evolution. From this transfer matrix we shall be able to get a Hamiltonian $H$ by taking its logarithm, i.e.,
$$U=U\left(z_\textrm{S},z_\textrm{T},z_\textrm{S}^*,z_\textrm{T}^* ,m_V,G_S,G_T\right)=e^{-\Delta t~H}\;.$$
In the next section we will define the function partition function $Z\left(z_\textrm{S},z_\textrm{T},z_\textrm{S}^*,z_\textrm{T}^*,m_V,G_S,G_T \right)$ and relate it with a one step evolution operator. The partition function is made of a bunch of tensors associated to the vertex, links and faces of the lattice, all of them contracted with each other as we will see next.

\subsection{The Partition Function}

The partition function of the model is defined on an oriented square lattice $\mathcal{L}$. We choose a square lattice for convenience. It can be easily defined on an arbitrarily triangulated lattice. The lattice is made up of vertices, links and faces glued together as shown in figure \ref{lattice1-b}. A tensor is associated for each vertex, link and face of $\mathcal{L}$ and contracted according to the gluing rules described below. In the next section we will show how the tensors we use to define the partition function are related with the structures constants of $\mathcal{A}$ and $H_n$.

\subsubsection{Associating Tensors to the Lattice}

 The procedure outlined helps define a function $\mathcal{L}\to \mathbb{C}$. 
 
 \begin{figure}[h!]
\centering
\subfigure[The tensor $M_{a_1 a_2 a_3 a_4}$ associated to a plaquette of the lattice.]{
\includegraphics[scale=1]{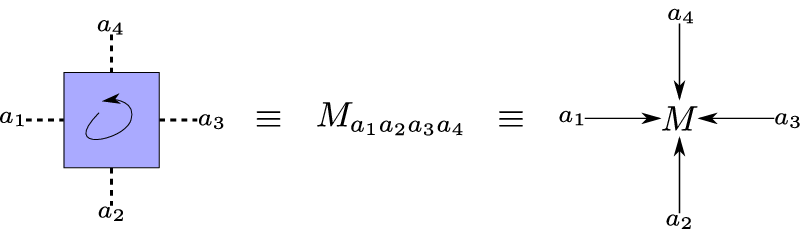} \label{tensorlattice-a}
}
\subfigure[The tensor $T^{\alpha_1 \alpha_2 \alpha_3 \alpha_4 \alpha_5 \alpha_6}$ associated to a vertex of the lattice.]{
\includegraphics[scale=1]{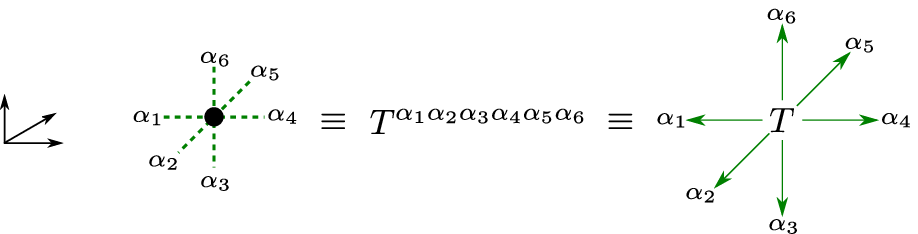} \label{tensorlattice-b}
}
\subfigure[The tensor ${L^{abcd\alpha}}_\beta$ associated to a link of the lattice.]{
\includegraphics[scale=1]{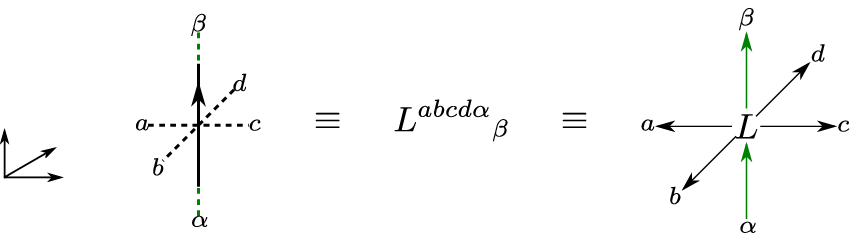} \label{tensorlattice-c}
}
\caption{The indices labelled by Latin letters (black arrows in the Kuperberg diagram) stand for the gauge fields while the indices labelled by Greek letters (green arrows in the Kuperberg diagram) stand for matter fields.}
\label{tensorlattice}
\end{figure} 
	A tensor $M_{a_1 a_2 a_3 a_4}$ is associated to each face. The four indices label the four links glued to this face. This is shown in figure \ref{tensorlattice-a}.  The indices are ordered counter clockwise according to orientation of the plaquette. Each vertex is glued into six different links and hence the tensor associated to it is denoted $T^{\alpha_1 \alpha_2 \alpha_3 \alpha_4 \alpha_5 \alpha_6}$ as shown in figure \ref{tensorlattice-b}. Finally each link is glued into four faces and two vertices, and so the tensor associated to a link is ${L^{abcd\alpha}}_\beta$ as shown in figure \ref{tensorlattice-c}. Note that the indices labelled by Latin letters (black arrows in the Kuperberg diagram) stand for the gauge fields while the indices labelled by Greek letters (green arrows in the Kuperberg diagram) stand for matter fields. In figure \ref{tensorlattice} the dotted lines mean that something is going to be glued to it.


%
%

\subsubsection{Contraction Rules}

\begin{figure}[h!]
\centering
\subfigure[The orientation tensor associated to the gauge part.]{
\includegraphics[scale=1]{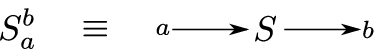} \label{orientationmap-a}
}
\hspace{2cm}
\subfigure[The orientation tensor associated to the matter part.]{
\includegraphics[scale=1]{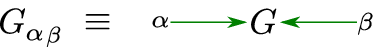} \label{orientationmap-b}
}
\caption{The tensors which take into account the orientation of the lattice.}
\label{orientationmap}
\end{figure}

To take care of the orientation of the lattice we still need two more tensors which will play the role of fixing orientation of the gauge part and the matter part. The one associated to the gauge part is the antipode map $S$ of the Hopf algebra $\mathcal{A}$ while the one associated to the matter part is some bilinear map $G:H_n \otimes H_n \to \mathbb{C}$ which we will describe later. The Kuperberg diagram for them are the ones shown in figure \ref{orientationmap}.

\begin{figure}[h!]
\centering
\subfigure[The contraction rule for the tensors in the gauge sector.]{
\includegraphics[scale=1]{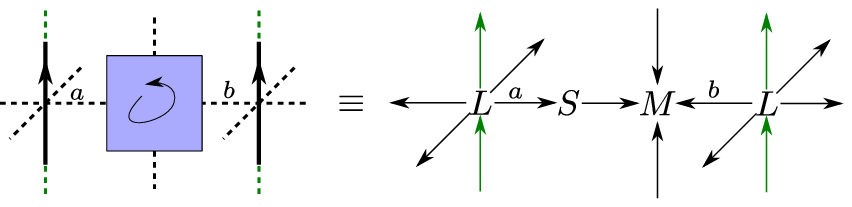} \label{contractionrule-a}
}
\hspace{1cm}
\subfigure[The contraction rule for the tensors in the matter sector.]{
\includegraphics[scale=1]{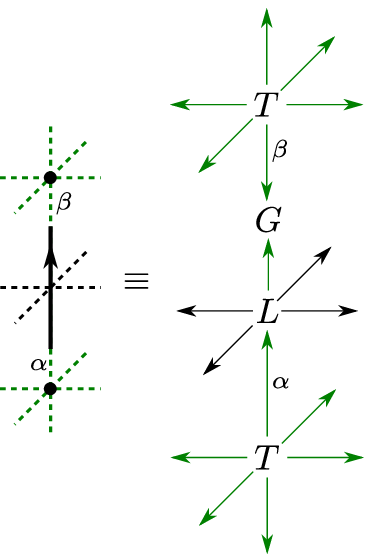} \label{contractionrule-b}
}
\caption{Contraction Rules.}
\label{contractionrule}
\end{figure}

The contraction tensors with the contraction rules are shown in figure \ref{contractionrule}. For the gauge part (contraction of a plaquette with a link) we use the antipode tensor when the orientation does not match or we contract the tensors $M_{a_1 a_2 a_3 a_4}$ with ${L^{abcd\alpha}}_\beta$ directly as shown in figure \ref{contractionrule-a}. For the matter part we contract the green arrow which is going out of the tensor ${L^{abcd\alpha}}_\beta$ using the tensor $G_{\alpha \beta}$ while the green arrow coming in the tensor ${L^{abcd\alpha}}_\beta$ we contract directly, as shown in figure \ref{contractionrule-b}

The partition function is defined as the contraction of all these tensors associated to the vertices, links and faces. Since the lattice we are considering has no boundary, $\partial \mathcal{L}=\emptyset$, there will be no free indices remaining after the gluing resulting in a scalar. As we will see in the next section the tensors $M_{a_1 a_2 a_3 a_4}$ and ${L^{abcd\alpha}}_\beta$ depend on the structure constants of $\mathcal{A}$, $H_n$, elements of the center of the algebra $\mathcal{A}$ and its dual, elements of $H_n$.  With these parameters the partition function can be written as  
$$Z\left(\mathcal{L}, z_\textrm{S},z_\textrm{T},z_\textrm{S}^*,z_\textrm{T}^*,m_V,G_S,G_T \right)=\prod_p M_{a_1 a_2 a_3 a_4}(p) \prod_v T^{\alpha_1 \alpha_2 \alpha_3 \alpha_4 \alpha_5 \alpha_6}(v)\prod_l {L^{abcd\alpha}}_\beta(l)\prod_o S_a^b \prod_o G_{\alpha \beta} $$
where the products runs over the orientations $(o)$, the plaquettes $(p)$, links $(l)$ and vertices $(v)$ of $\mathcal{L}$.

\subsection{The Transfer Matrix $U$ from the Partition Function $Z$}

\begin{figure}[h!]
\begin{center}
		\includegraphics[scale=1]{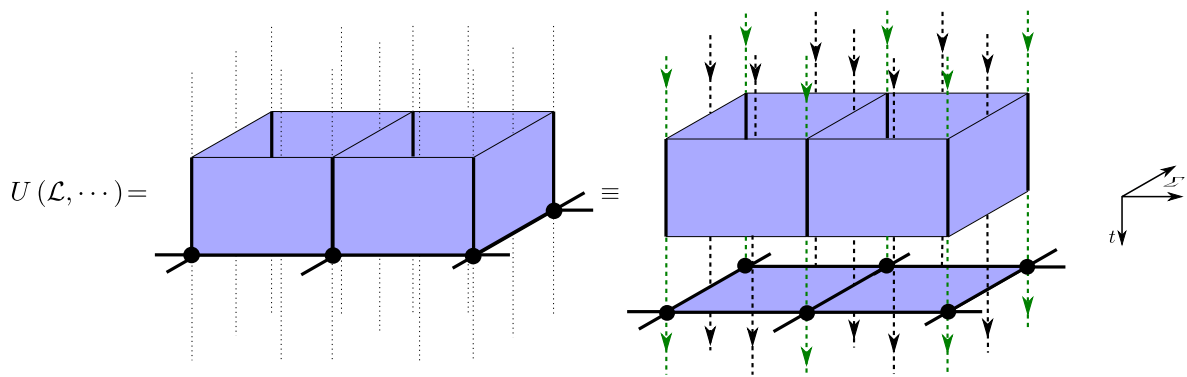}
	\caption{The one step evolution operator $U$.}
\label{onestepevolution}
\end{center}
\end{figure}

Having defined the partition function we can now define a one step evolution operator $U$ such that 
$$Z=\textrm{tr}(U^N)\;.$$
The way to do this is by looking at the lattice in which the partition function is defined on and to take slices of it in the time direction. We now introduce a new point of view where we look at the lattice as a contraction of a bunch of tensors as defined above. As we will see sometimes it will be convenient to think of the lattice as a tensor network and sometimes more convenient to think of it as a gluing of vertices, links and faces.
\begin{figure}[h!]
\begin{center}
		\includegraphics[scale=1]{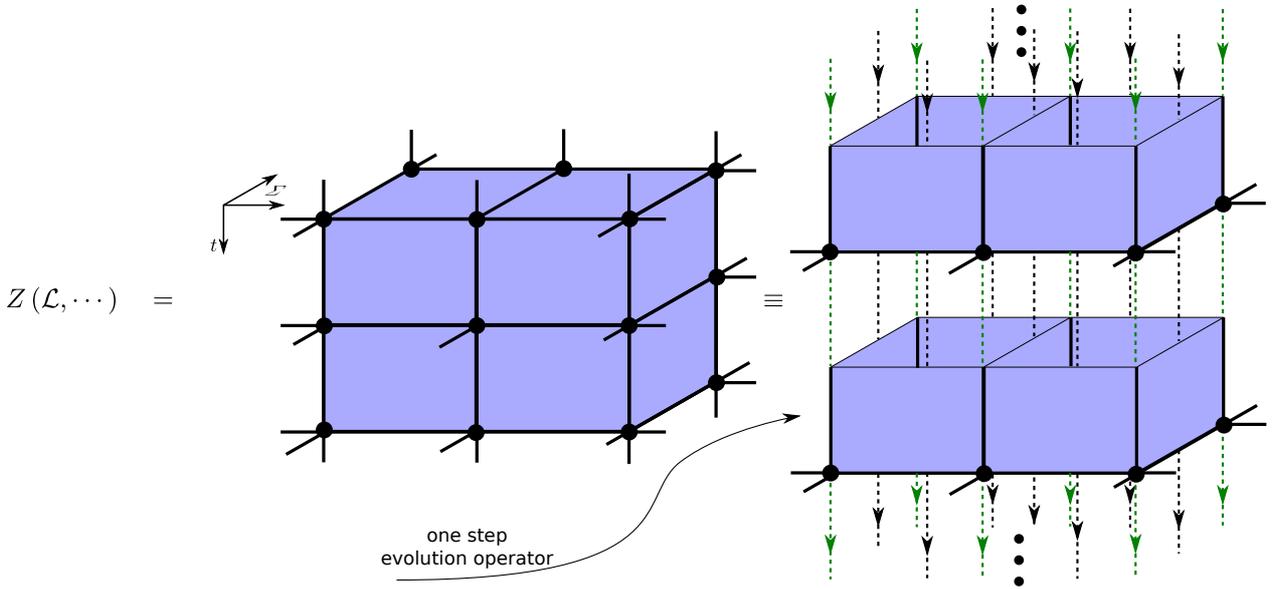}
	\caption{The partition function as the trace of a product of a one step evolution operator.}
\label{transfermatrix}
\end{center}
\end{figure}
In figure \ref{transfermatrix} the black and green arrows represent the free indices (gauge and matter degrees of freedom) of the one step evolution operator. The one step evolution operator is the one shown in figure \ref{onestepevolution}, it is made of a bunch of boxes without their caps glued one besides the other.   

\section{The Transfer Matrix as a Product of Local Operators}

We will see that the one step evolution defined in figure \ref{onestepevolution} is a product of local operators. These local operators act on vertices, links and plaquettes of the lattice, and hence are called vertex, link and plaquette operators, respectively. But before we split it we must have a better definition of the tensors used to build it. In other words we need to know how they are related with the structure constants of $\mathcal{A}$ and $H_n$. In the next section we will define their algebraic structure which is a very important part of the model.

\subsection{The Structure Constants}

\begin{figure}[h!]
\centering
\subfigure[The multiplication map.]{
\includegraphics[scale=1]{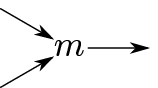} \label{hopfstructureconstants-a}
}
\hspace{3cm}
\subfigure[The co-product map.]{
\includegraphics[scale=1]{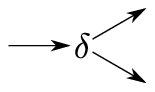} \label{hopfstructureconstants-b}
}
\hspace{3cm}
\subfigure[The antipode map.]{
\includegraphics[scale=1]{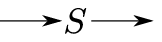} \label{hopfstructureconstants-c}
}
\caption{Structure constants of the involutory Hopf algebra $\mathcal{A}$.}
\label{hopfstructureconstants}
\end{figure}

Let $\mathcal{A}$ be an involutory Hopf algebra with a product $m:\mathcal{A}\otimes \mathcal{A} \to \mathcal{A}$, a co-product $\delta:\mathcal{A} \to \mathcal{A} \otimes \mathcal{A}$ and a involutory map $S:\mathcal{A} \to \mathcal{A}$. Its structure constants are the tensors shown in figure \ref{hopfstructureconstants}.

\begin{figure}[h!]
\centering
\subfigure[The action map of $\mathcal{A}$ on $H_n$.]{
\includegraphics[scale=1]{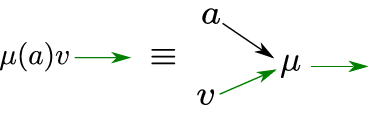} \label{modulestructureconstants-a}
}
\hspace{1cm}
\subfigure[The co-structure map in $H_n$.]{
\includegraphics[scale=1]{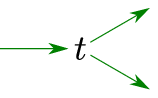} \label{modulestructureconstants-b}
}
\hspace{1cm}
\subfigure[The algebra action on the module is a homomorphism.]{
\includegraphics[scale=1]{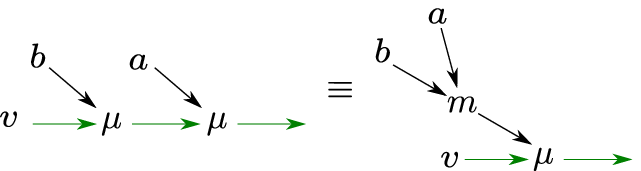} \label{modulestructureconstants-c}
}
\caption{Structure constants involving the $H_n$ space.}
\label{modulestructureconstants}
\end{figure}

Let $H_n$ be a module of $\mathcal{A}$ which has a semi-simple co-algebra structure $t: H_n\to H_n \otimes H_n$. We define the action of $\mathcal{A}$ on $H_n$ by the map $\mu: \mathcal{A}\otimes H_n \to H_n$ in the following way
$$\mu:a\otimes v\mapsto\mu(a) \rhd v\;.$$
The structure constants involving $H_n$ are the ones shown in figure \ref{modulestructureconstants-a} and \ref{modulestructureconstants-b}. We also show the action of the algebra $\mathcal{A}$ on $H_n$ with respect to the product of the algebra $\mathcal{A}$, in other words we want this action to be a homomorphism, i. e., $\mu(ab)\rhd v = \mu(a)\rhd\left(\mu(b)\rhd v \right)\; \forall v$ as shown in figure \ref{modulestructureconstants-c}.

\subsection{Building the Tensors}

\begin{figure}[h!]
\begin{center}
		\includegraphics[scale=1]{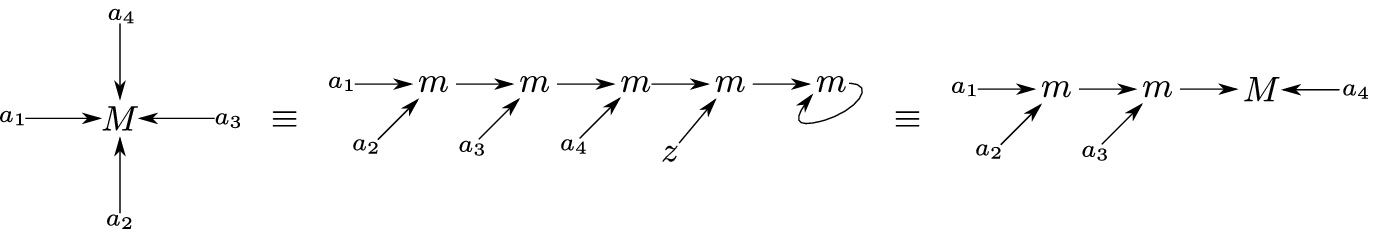}
	\caption{The definition of the tensor $M_{a_1 a_2 a_3 a_4}$.}
\label{tensordefinition-a}
\end{center}
\end{figure}

The first tensor we will define is the tensor $M_{a_1 a_2 a_3 a_4}$. Its definition is shown in figure \ref{tensordefinition-a}. Due to the associativity of the algebra the tensor $M_{a_1 a_2 a_3 a_4}$ is invariant under cyclic permutation of its indices and can be written in different ways as shown in figure \ref{tensordefinition-a}. The element $z$ is an element of the center of the algebra.

\begin{figure}[h!]
\begin{center}
		\includegraphics[scale=1]{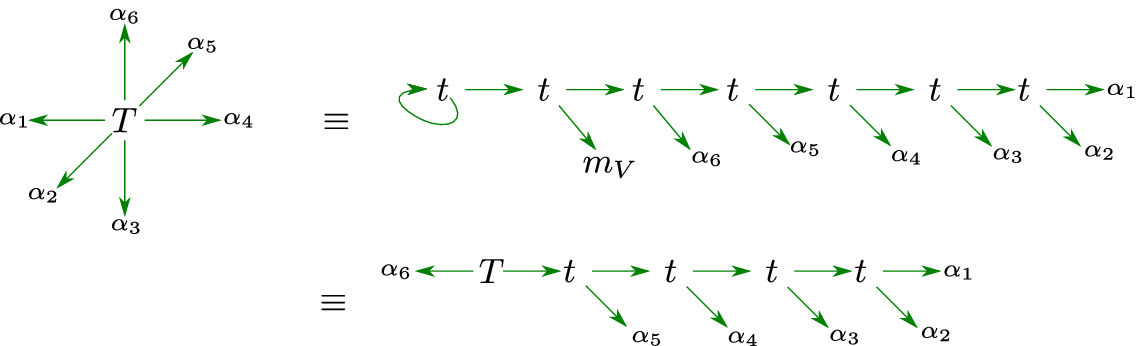}
	\caption{The definition of the tensor $T^{\alpha_1 \alpha_2 \alpha_3 \alpha_4 \alpha_5 \alpha_6}$.}
\label{tensordefinition-b}
\end{center}
\end{figure}

The tensor $T^{\alpha_1 \alpha_2 \alpha_3 \alpha_4 \alpha_5 \alpha_6}$ is made of the co-structure map in $H_n$ as shown in figure \ref{tensordefinition-b}. As the co-product is co-commutative this tensor is invariant under interchange of a pair of its indices.  Like the tensor $M_{a_1 a_2 a_3 a_4}$ this tensor can also be written in different equivalent forms as shown in figure \ref{tensordefinition-b}.

\begin{figure}[h!]
\begin{center}
		\includegraphics[scale=1]{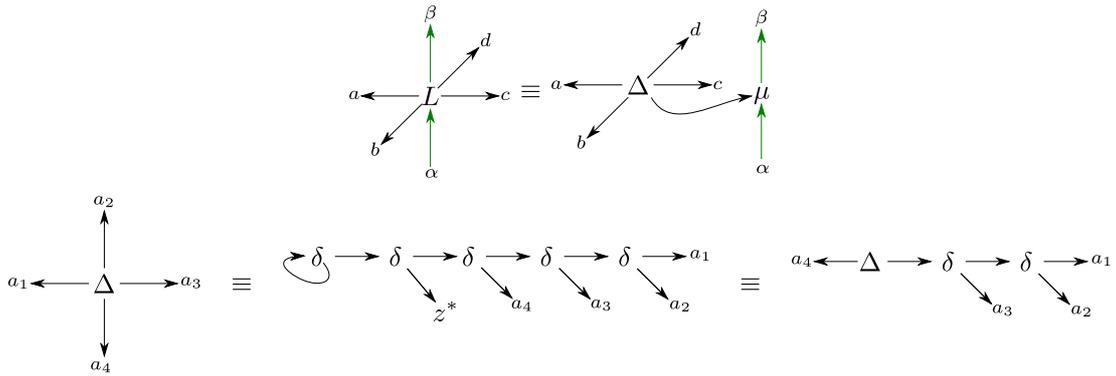}
	\caption{The definition of the tensor ${L^{abcd\alpha}}_\beta$.}
\label{tensordefinition-c}
\end{center}
\end{figure}

Finally the tensor associated to the links given by ${L^{abcd\alpha}}_\beta$ is defined in figure \ref{tensordefinition-c}. As we can see it is made up of the algebra action on $H_n$ given by the $\mu$ map and the tensor $\Delta^{a_1 a_2 a_3 a_4 a_5}$ which is also defined in figure \ref{tensordefinition-c}. The element $z^*$ is an element of the co-center of the algebra $\mathcal{A}^*$.

\subsection{Splitting the One Step Evolution Operator}\label{sec:split}

\begin{figure}[h!]
\begin{center}
		\includegraphics[scale=1]{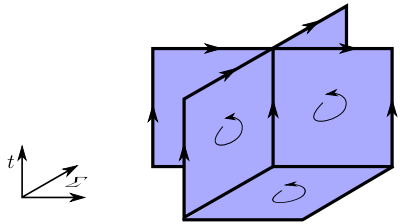}
	\caption{The orientation of the lattice considered.}
\label{fixingorientation}
\end{center}
\end{figure}

In order to make the procedure clear we will represent the spacelike part as being made of spacelike plaquettes and spacelike links and the timelike part as being made of timelike plaquettes and timelike links. The splitting will be done in two steps: first we will split the tensors associated to spacelike links and vertices and then we split the tensor associated to the timelike plaquettes. These two steps show how the one step evolution operator can be written as a product of local operators. Although the model does not depend on the orientation of the lattice we have to set up some orientation in order to use the contraction rules described above. So we consider the orientation shown in figure \ref{fixingorientation}.

\subsubsection{Splitting the Spacelike Part of the Tensor Network}

\begin{figure}[h!]
\begin{center}
		\includegraphics[scale=1]{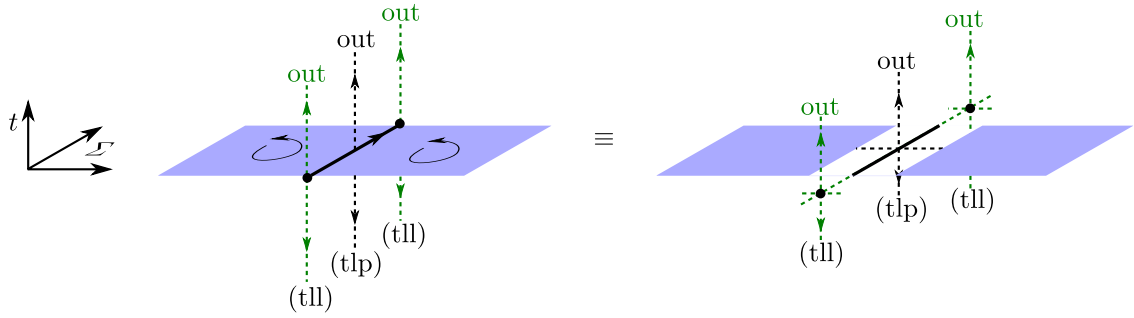}
	\caption{A spacelike link shared by two adjacent plaquettes and two vertices.}
\label{splittingSLL-a}
\end{center}
\end{figure}

Consider a spacelike link which is shared by two adjacent plaquettes and two vertices as shown in figure \ref{splittingSLL-a}.

\begin{figure}[h!]
\begin{center}
		\includegraphics[scale=1]{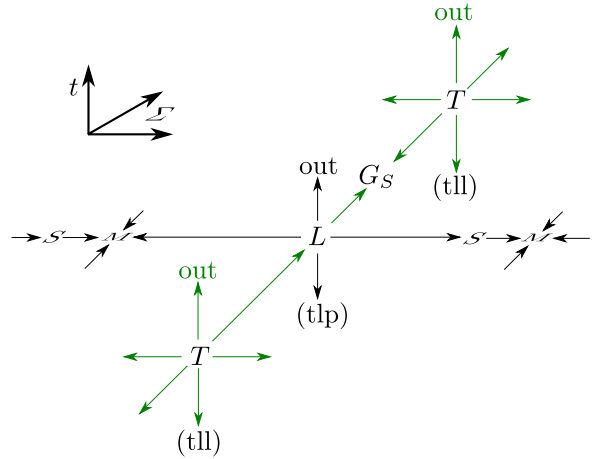}
	\caption{Tensor network associated to the picture in figure \ref{splittingSLL-a}.}
\label{splittingSLL-b}
\end{center}
\end{figure}

The tensor network associated to the figure \ref{splittingSLL-a} is shown in figure \ref{splittingSLL-b}. In the picture tlp and tll stand for timelike plaquette and timelike link respectively.

\begin{figure}[h!]
\begin{center}
		\includegraphics[scale=1]{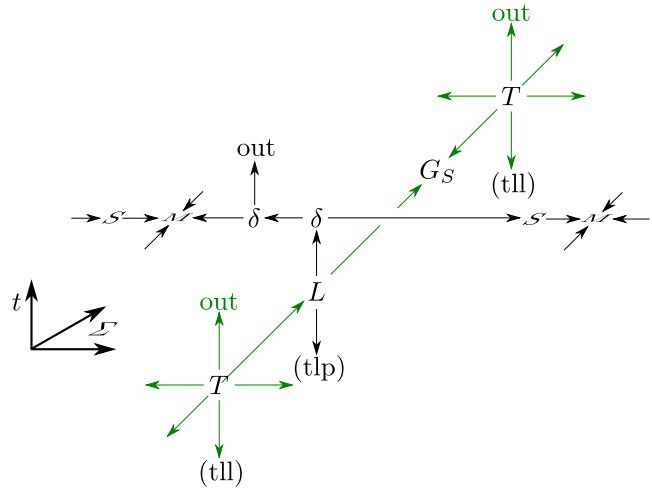}
	\caption{Tensor network associated to the picture in figure \ref{splittingSLL-a} with the tensor ${L^{abcd\alpha}}_\beta$ splitted.}
\label{splittingSLL-c}
\end{center}
\end{figure}

As seen in figure \ref{tensordefinition-c} the tensor ${L^{abcd\alpha}}_\beta$ can be splitted in terms of the structure constants representing the co-product of the algebra. Thus the diagram in figure \ref{splittingSLL-b} reduces to the one shown in figure \ref{splittingSLL-c}.

\begin{figure}[h!]
\begin{center}
		\includegraphics[scale=1]{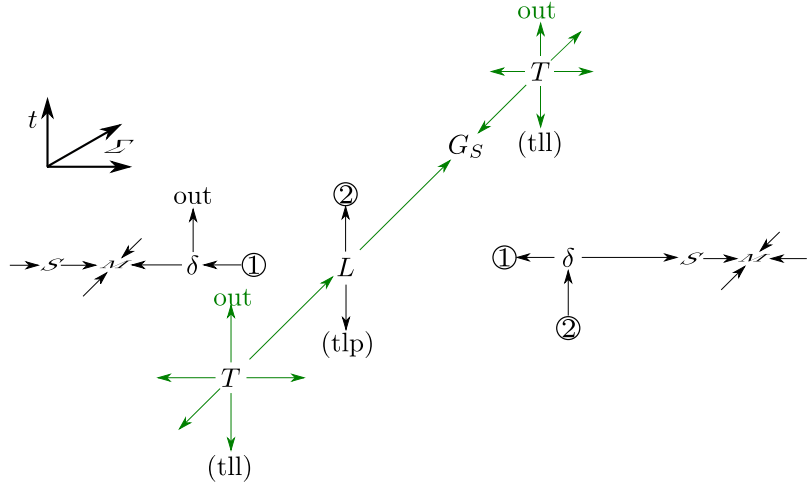}
	\caption{The tensors network in figure \ref{splittingSLL-c} rewritten for clarity}
\label{splittingSLL-d}
\end{center}
\end{figure}

To make things clear we write down the same tensor network of figure \ref{splittingSLL-c} in figure \ref{splittingSLL-d}, where we have separated the tensors associated to each one of the spacelike plaquettes. Each leg of the tensor $M_{a_1 a_2 a_3 a_4}$ will be contracted to a tensor $\delta_a^{cd}$ (directly or indirectly by the antipode map) and that will lead to the plaquette operator as we shall soon see.
 
\begin{figure}[h!]
\begin{center}
		\includegraphics[scale=1]{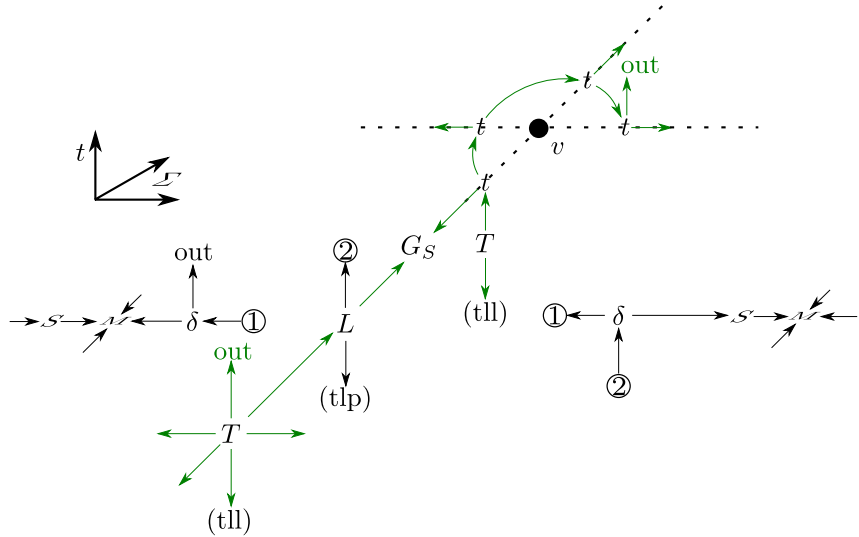}
	\caption{Splitting the tensors associated to a vertex.}
\label{splittingSLL-e}
\end{center}
\end{figure}

We now split the tensor associated to each vertex in figure \ref{splittingSLL-d}. For that we just use the definition of the tensor $T^{\alpha_1 \alpha_2 \alpha_3 \alpha_4 \alpha_5 \alpha_6}$ in figure \ref{tensordefinition-b}. The splitting shown in figure \ref{splittingSLL-e} has to be done for the all the vertices of the lattice, but here for simplicity we are illustrating just for one single vertex. 

\begin{figure}[h!]
\centering
\subfigure[The plaquette operator as a tensor network.]{
\includegraphics[scale=1]{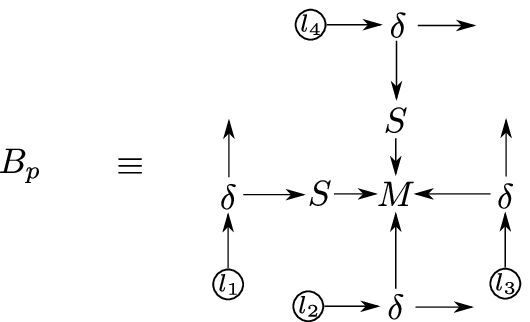} \label{splittingSSL-f}
}
\hspace{1cm}
\subfigure[The plaquette $p$ where the operator $B_p$ acts on.]{
\includegraphics[scale=1]{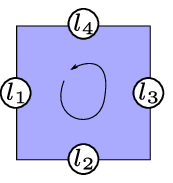} \label{splittingSSL-ff}
}
\hspace{1cm}
\subfigure[The tensor network which will be contract with the timelike part of the remaining tensor network.]{
\includegraphics[scale=1]{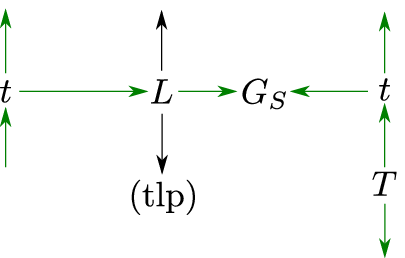} \label{splittingSSL-g}
}
\caption{The spacelike part of the tensor network can be written as a product of these two blocks of tensor networks.}
\label{splittingSSL-fg}
\end{figure}

The tensor network in figure \ref{splittingSLL-d} is rewritten in a separated way as shown in figure \ref{splittingSLL-e}. After all this splitting we see that the spacelike tensor network part can be written as a product of two blocks of tensor networks which are the ones shown in figure \ref{splittingSSL-fg}. The first one (figure \ref{splittingSSL-f} is the operator called the plaquette operator. The second one in figure \ref{splittingSSL-g} is called $K_l$ and it will be contracted with the timelike tensor network part to build the others operators which make up the transfer matrix.
 
\subsubsection{Splitting the Timelike Part of the Tensor Network}

\begin{figure}[h!]
\centering
\subfigure[A timelike plaquette of the transfer matrix.]{
\includegraphics[scale=1]{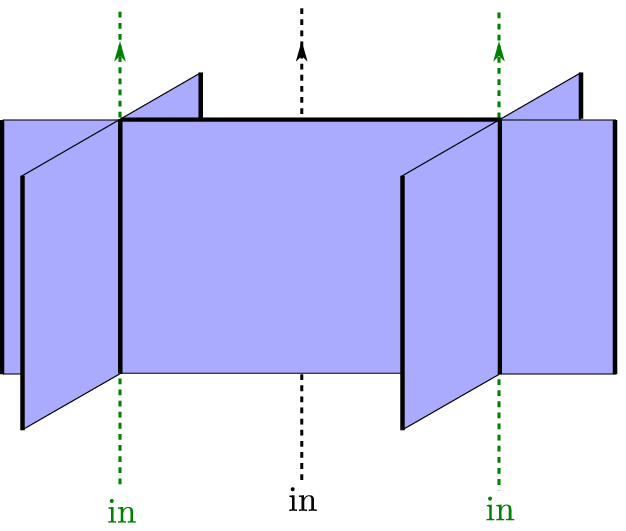} \label{splittingTLP-a}
}
\hspace{2cm}
\subfigure[A timelike plaquette with the tensor network associated to the spacelike link attached above.]{
\includegraphics[scale=1]{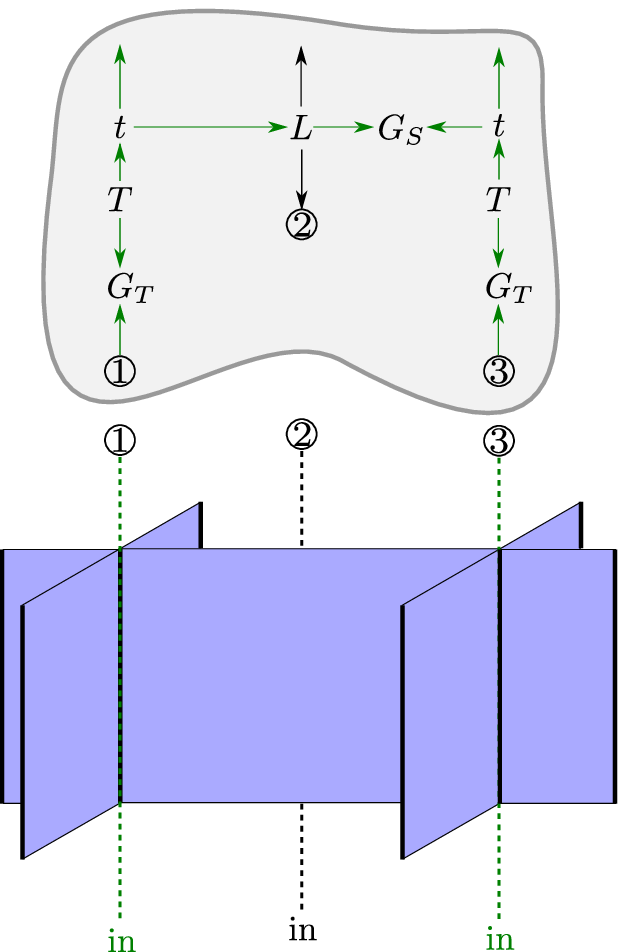} \label{splittingTLP-b}
}
\caption{Timelike part of the transfer matrix.}
\label{splittingTLP-ab}
\end{figure}

We now split the timelike part of the tensor network. For that consider a timelike plaquette which is shared by two timelike links and one spacelike link, as shown in the figure \ref{splittingTLP-a}. The spacelike link on the top is the one which has already been splitted and it led to the tensor network attached to the timelike plaquette as shown in figure \ref{splittingTLP-b}.

\begin{figure}[h!]
\centering
\subfigure[The tensor network associated to the timelike part of the diagram.]{
\includegraphics[scale=1]{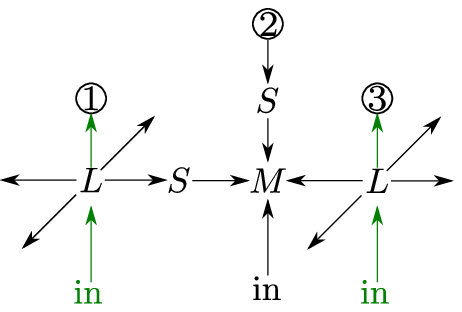} \label{splittingTLP-c}
}
\hspace{2cm}
\subfigure[The tensor $M_{a_1 a_2 a_3 a_4}$ splitted in terms of the structure constants of the algebra.]{
\includegraphics[scale=1]{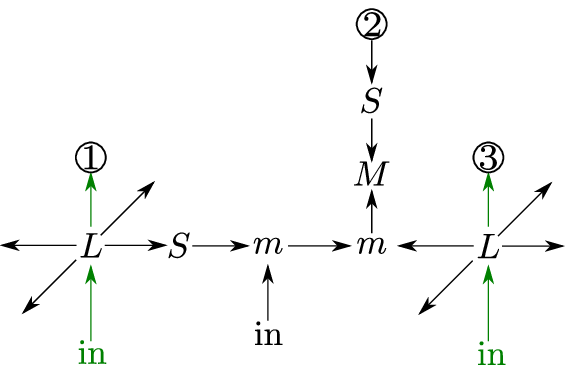} \label{splittingTLP-d}
}
\caption{Splitting the tensor network of the timelike part of the transfer matrix.}
\label{splittingTLP-cd}
\end{figure}

The tensor network associated to the timelike plaquette contracted with the two timelike links is the one one shown in figure \ref{splittingTLP-c} and it can be easily changed to the one in figure \ref{splittingTLP-d} by using the definition of the tensor $M_{a_1 a_2 a_3 a_4}$ in figure \ref{tensordefinition-a}.

\begin{figure}[h!]
\begin{center}
		\includegraphics[scale=1]{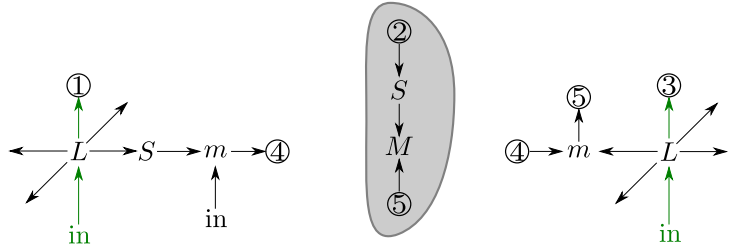}
	\caption{Rewriting the tensor network in figure \ref{splittingTLP-d} with the tensors separated.}
\label{splittingTLP-e}
\end{center}
\end{figure}

As before we rewrite the tensor network in figure \ref{splittingTLP-d} as the one in figure \ref{splittingTLP-e}.

\begin{figure}[h!]
\centering
\subfigure[The vertex operator.]{
\includegraphics[scale=1]{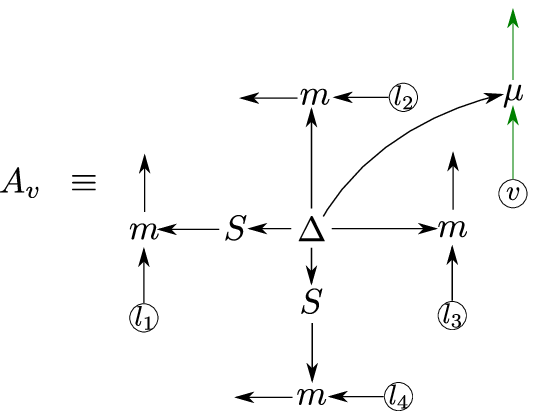} \label{splittingTLP-f}
}
\hspace{1cm}
\subfigure[The vertex $v$ where the vertex operator $A_v$ acts.]{
\includegraphics[scale=1]{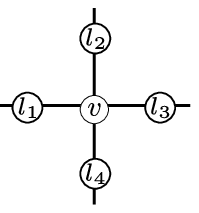} \label{splittingTLP-ff}
}
\hspace{1cm}
\subfigure[The tensor network of the link operator.]{
\includegraphics[scale=1]{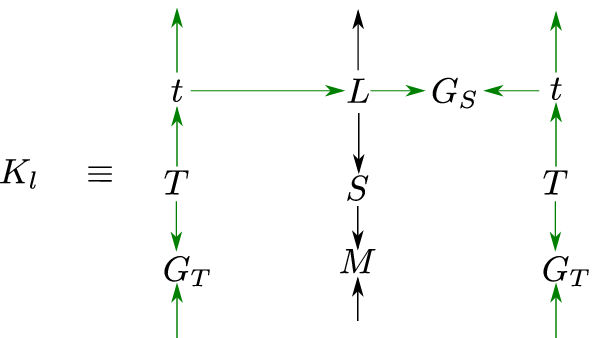} \label{splittingTLP-g}
}
\caption{Vertex operator and the operator $K_l$ which acts on a link.}
\label{splittingTLP-fg}
\end{figure}

In order to keep it clear in our mind where are these tensors acting on one should take a look at the figure \ref{splittingTLP-b}. Note that each and every timelike tensor ${L^{abcd\alpha}}_\beta$ will have a tensor $m_{ab}^c$ contract with each of its gauge legs (directly or indirectly with the antipode). It lead to an operator $A_v$ for each timelike link of the lattice, or in other words, for each vertex spacelike vertex $v$. This operator is the one drawn in figure \ref{splittingTLP-f}. The tensor in highlight in figure \ref{splittingTLP-e} is now attached to the tensor network on the top in figure \ref{splittingTLP-b} and it give us a new operator $K_l$ which acts on the links, see figure \ref{splittingTLP-g}


The operator $K_l$ is the same on which appears in the (1+1) dimensional models. In the appendix we show how the transfer matrices of such a models can be obtained e also that the $K_l$ operator can be written in terms of simplers operators. The transfer matrix we started with can now be written as
$$U\left(z_\textrm{S},z_\textrm{T},z_\textrm{S}^*,z_\textrm{T}^*,m_V,G_S,G_T \right)=\prod_p B_p(z_S) \prod_l K_l \prod_v A_v(z_T^*)\;.$$
But the $K_l$ operator, as can be seen in the appendix, can also be written as 
$$\prod_l K_l = \prod_l C_l(G_S)  \prod_l\left(T_l(z_S^*)D_l L_l(z_T) \right)\prod_v\tilde{V}_v(G_T) Q_v(m_v)\;,$$
where the operators $Q_v(m_v)$, $T_l(z_S^*)$ and $L_l(z_T)$ will be explained in the next section. The operators $D_l$ and $\tilde{V}_v(G_T)$ are explained in the appendix. The operator $C_l(G_S)$ is called the link operator and acts on a link and on the its vertex. This operator is defined below in figure \ref{splittingTLP-h}.  
\begin{figure}[h!]
\centering
\subfigure[The link operator.]{
\includegraphics[scale=1]{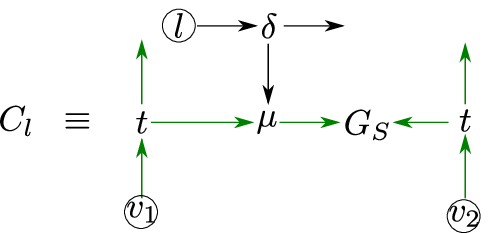} \label{splittingTLP-h}
}
\hspace{2cm}
\subfigure[The link $l$ where the link operator $C_l$ acts.]{
\includegraphics[scale=1]{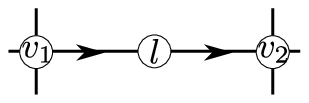} \label{splittingTLP-hh}
}
\caption{The link operator.}
\label{splittingTLP-hh}
\end{figure}

\section{The Transfer Matrix $U$ in its Final Form}

So far we have used a diagrammatic language where we associated tensors to the vertices, links and plaquettes of a triangulated three manifold to build a tensor network and showed that this results in a partition function for a closed manifold and a transfer matrix for a manifold with boundaries. We then used this transfer matrix represented by a tensor network to find out the local operators which can be used to piece up such a network. In this section we will write down the algebraic expressions for these operators and study their properties. For simplicity we choose the gauge algebra to belong to groups and so the states on the links are elements of the group algebra of a gauge group denoted by $\mathbb{C}(G)$. The matter degrees of freedom belong to a module of $\mathbb{C}(G)$ which we denote by $H_n$, in other words $H_n$ is a vector space carrying an action of $\mathbb{C}(G)$. The formalism developed so far can be applied to any involutory {\bf $C^*$}-Hopf algebra and its corresponding module. 

The operators that make up the transfer matrix can be divided into the ones which act only on the gauge fields, the ones that only act on the matter fields and the ones that involve the coupling between the gauge and the matter fields through the $\mu$ map defined in figure (\ref{tensordefinition-c}). We describe each set separately before we write down the full transfer matrix.

\begin{figure}[h!]
\centering
\subfigure[left multiplication.]{
\includegraphics[scale=1]{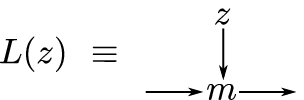} \label{LRTQops-a}
}
\hspace{.5cm}
\subfigure[right multiplication.]{
\includegraphics[scale=1]{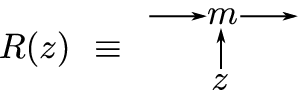} \label{LRTQops-b}
}
\hspace{.5cm}
\subfigure[gauge projector.]{
\includegraphics[scale=1]{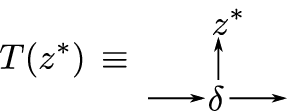} \label{LRTQops-c}
}
\hspace{.5cm}
\subfigure[matter projector.]{
\includegraphics[scale=1]{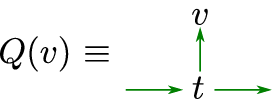} \label{LRTQops-d}
}
\caption{These operators are the ones which we use to build the vertex, plaquette and link operators.}
\label{LRTQops}
\end{figure}

In order to write an algebraic expression for the operators we derived in previous section we need to define three operators which are made of the structure constants of the algebra and the module, as shown in figure \ref{LRTQops}.
These operators are linear on its parameters, in other words, $L(z)=\sum_g z^gL(\phi_g)$ (the same for $R(z)$, $T(z^*)$ and $Q(v)$) and they act on the vector basis as
\begin{eqnarray}
L(\phi_g)\vert h \rangle &=& \vert gh\rangle \;, \nn \\
R(\phi_g)\vert h \rangle &=& \vert hg \rangle \;, \nn \\ 
T(\Psi_g)\vert h \rangle &=& \delta(g,h) \vert h\rangle \nn \;,\\
Q(\chi_i)\vert j \rangle &=& \delta(i,j) \vert j\rangle  \label{lrtopss}\;.
\end{eqnarray}
where we have used the bra-ket notation for elements of the basis ($\vert g \rangle := \phi_g$ and $\vert i \rangle = \chi_i$). Sometimes we use the short notation $L^g=L(\phi_g)$, $R^g=R(\phi_g)$, $T^g=T(\Psi^g)$ and $Q^i=Q(\chi_i)$. Now using  this operators in the next section we write the plaquette, vertex and link operators derived before in terms of them. There are three operators acting on the gauge sector alone. These are given by the plaquette operator and two operators acting on the qudits on the links. They can be deduced from their respective tensor network representation from the previous sections. We write down each of these in what follows. We will also see how the different parameters parametrizing the transfer matrix $U\left(z_S, z_T, z_S^*, z_T^*, m_V, G_S,G_T\right)$ arise in the definition of these operators. This will also show how the transfer matrix depends on these parameters. 

\subsection{The Gauge Sector}

The plaquette operator can be written down by looking at its tensor network representation given in figure (\ref{splittingSSL-f}). We find

\beq\label{algBp} B_p  =  \sum_{C\in [G]} \beta_C B_p^C \eeq
where $C$ labels a particular conjugacy class from the set of conjugacy classes in $G$ denoted by $[G]$. Each of $B_p^C$ is given by
\beq\label{algBpc}  B_p^C =\vert G \vert  \sum_{g\in C}\sum_{\{g_i\}}\delta(g_1g_2g_3g_4 g,1_G) T^{g_1^{-1}}_{i_1}\otimes T^{g_2}_{i_2}\otimes T^{g_3}_{i_3}\otimes T^{g_4^{-1}}_{i_4} \eeq
where $1_G$ is the identity of the group $G$.

While writing down the plaquette operator in Eq.(\ref{algBp}) we have made use of the fact that the tensor associated to the spacelike plaquette $M_{a_1a_2a_3a_4}$ contains a central element of $\mathbb{C}(G)$ as can also be seen in figure (\ref{tensordefinition-a}). For the plaquette operator in Eq.(\ref{algBp}) this central element $z_S$, where $S$ denotes spacelike, is given by
\beq z_S = \sum_{C\in [G]}\beta_C\sum_{g\in C} \phi_g \eeq
with the set $\{\phi_g\}$ forming a basis of $\mathbb{C}(G)$.The action of the plaquette operator on the oriented square lattice is shown in figure \ref{splittingSSL-ff}.

The operators $B_p^C$ form a complete basis of orthogonal projectors since
$$B_p^C B_p^{C^\prime}=\delta(C,C^\prime)B_p^C\;,$$
and also
$$\sum_C B_p^C = \mathbb{I}\;.$$

One of the link operators acting on the qudit on the links is given by 
\beq\label{algZ} T_i(z_S^*) = \sum_{g\in G} b_g T^g_i \eeq
where $b_g$ are constants and $T^g_i$ is gauge projector defined in figure \ref{LRTQops-c}. These operators arise by the parameter on spacelike links given by $z_S^*$ which is a central element of the dual of $\mathbb{C}(G)$. They can also be seen in figure (\ref{tensordefinition-c}).  This central element is given by
\beq z_S^* = \sum_{g\in G} b_g\psi_g \eeq
with the set $\{\psi_g\}$ forming the basis of the dual of $\mathbb{C}(G)$. 

The second link operator acting in the gauge sector is given by
\beq\label{algX} L_j(z_T) = \sum_{R\in \textrm{IRR's of}~ G}a_R L_j(z_R) \eeq
where $a_R$'s are constants and $z_R$ is given by
\beq \label{algXr} z_R  = \frac{1}{|G_R|}\sum_{g\in G}\chi_R(g)\phi_g. \eeq
$\chi_R(g)$ is the character of the element $g$ in the IRR $R$ and $|G_R|$ is the number of elements with non-zero trace in the IRR $R$. The parameters $z_T$ is given by
\beq\label{algL} z_T=\sum_R a_R z_R \eeq 
We could as well have defined the operator $R_j(z_T)$, but since $z_T$ is an element of the center of the algebra these two operators are the same, namely $L_j(z_T)=R_j(z_T)$.

The operator $X_j$ in Eq.(\ref{algX}) is obtained through the parameter $z_T$ in the timelike plaquette. This is an element of the center of $\mathbb{C}(G)$ given by
\beq z_T = \sum_R a_R\sum_{g\in G}\chi_R(g)\phi_g. \eeq

These operators exhaust the operators acting only in the gauge sector. We now turn to those acting only in the matter sector.

\subsection{The Matter Sector}

There are two operators which act only in the matter sector. One of them is obtained by parametrizing the vertices by $m_V$, an element of the module $H_n$ and the other by parametrizing the timelike links with the inner product $G_T$. The operator parametrized by $m_V$ is given by
\beq\label{algQ} Q_v(m_v) = \sum_{i=1}^n c_i Q_v^{i} \eeq
where $c_i$ are constants and $Q_v^{i}$ is given by
\beq \label{actQ}Q_v^{i} |j\rangle = \delta_{i, j}|j\rangle \eeq
where $\chi_{i,j}$ are elements of the module $H_n$. For the definition of $Q_v$ in Eq.(\ref{algQ}) the element $m_V$ is given by
\beq m_V  = \sum_{i=1}^n c_i \chi_i.\eeq

As the module only has a co-structure this operator is very similar to the coproduct map in the gauge sector. In fact for the module $H_n$ with symmetry group $\mathbb{Z}_n$ we can identify $Q_v^{j}$ with the $T^{\omega^j}$ operator of the gauge sector for $\mathbb{C}(\mathbb{Z}_n)$ with $\omega=e^{\frac{2\pi i}{n}}$ and $j\in \left(0,\cdots, n-1\right)$. If we label the elements of the group $\mathbb{Z}_n$ as $\{\omega^k;~ k\in\left(0,\cdots, n-1\right)\}$ we can then write $Q_v^{l}$ as
\beq \label{algQa}Q_v^{l} = \frac{1}{n}\sum_{k=0}^{n-1}\chi_{\omega^l}(\omega^k) Z_v^k \eeq
where $\chi_{\omega^l}(\omega^k)$ is the character of the element $\omega^k$ in the IRR of $\mathbb{Z}_n$ labelled by $\omega^l$. There are $n$ such expressions corresponding to the $n$ IRR's of $\mathbb{Z}_n$. 

The operator $Z_v^k$ in Eq.(\ref{algQa}) is a generator of $\mathbb{Z}_n$ and is defined by
\beq Z_v|\omega^k\rangle = \omega^k|\omega^k\rangle. \eeq 

We can also consider modules $H_n$ with other symmetry groups, especially non-Abelian groups, and in this case the expression for $Q_v^{i}$ is different from the one given by Eq.(\ref{algQa}). The symmetry groups of the module $H_n$ can also be thought of as global symmetry groups of the system. 

We now consider the operators acting on both the gauge and matter sectors. 

\subsection{The Gauge + Matter Sector}

There are two operators acting on both the gauge and matter sector. They are the operators coupling the two sectors. These operators are the vertex and link operators given in figures (\ref{splittingTLP-f}) and (\ref{splittingTLP-h}) respectively.

The vertex operator $A_v$ is given by 
\beq\label{algAv} A_v = \sum_{g\in G}\alpha_g A_v^g \eeq
where 
\beq  A_v^g = \mu_v(\phi_g)\otimes L_{j_1}^g\otimes L_{j_2}^{g}\otimes R_{j_3}^{g^{-1}}\otimes R_{j_4}^{g^{-1}} \eeq
with $\mu_v(\phi_g)$ being the representation of the gauge group on the matter field located on the vertex $v$. The single qudit operators $L^g_i$ and $R^g_i$, acting on the gauge fields located on the links, are given by Eq.(\ref{lrtopss}).  

The vertex operator $A_v$ is obtained by using the parameter $z_T^*$ in the transfer matrix given by
\beq z_T^* = \sum_{g\in G} \alpha_g\psi_g.\eeq
This is a parameter living on the timelike links of the $2+1$ dimensional manifold. 

For particular choices of the parameter $z_T^*$ we obtain the set of orthogonal vertex operators which are projectors. These parameters are given by
\beq z_T^* = \frac{1}{|G_R|} \sum_{g\in G} \chi_R(g) \psi_g \eeq for the different $R$'s in the set of IRR's of $G$. 
They result in the following set of vertex operators 
\beq A_v^R  = \frac{1}{|G_R|} \sum_{g\in G} \chi_R(g) A_v^g.\eeq

This vertex operator is very similar to the one defined for the quantum double models~\cite{pp1, Kit, Agu} with the addition of the representation of the gauge group acting on the vertex part. Thus it can still be thought of as a gauge transformation just as in the case of the quantum double models.

The other operator acting on both the gauge and the matter sector is the link operator represented as a tensor network in figure (\ref{splittingTLP-h}). In terms of operators this operator is given by 
\beq\label{algC} C_l = \sum_{\chi_{v_1},\;\chi_{v_2},\;\phi_l} G_S(\mu(\phi_l)\chi_{v_1},\chi_{v_2})Q_{v_1}^{\chi_{v_1}} T_l^{\phi_l} Q_{v_2}^{\chi_{v_2}} \eeq 
where the matrix $G_S$ implements the inner product between the vectors in the module $H_n$. 

\begin{figure}[h!]
\begin{center}
		\includegraphics[scale=1]{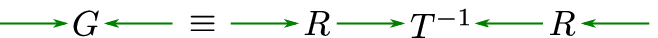}
	\caption{The tensor network representation for the inner product $G$ in terms of the matrix $R$.}
\label{pc8}
\end{center}
\end{figure}

This inner product $G$ can also be thought of to be represented by the following tensor network shown in figure (\ref{pc8}), 
where $T^{-1}$ is a tensor such that $(T^{-1})_{\alpha\beta}T^{\beta \gamma} = \delta(\alpha,\gamma)$, in other words, $(T^{-1})_{\alpha\beta}= \delta(\alpha,\beta)$.

The matrices $R$ in this definition can be thought of as those performing basis transformations on the elements of the module $H_n$. If the matrix $G$ is chosen to be identity, it is the same as choosing $R$ to be identity. For other choices of $R$ we can find the operators orthogonal to $C_l$. This will soon become clearer when we illustrate using examples. 

We thus have the operators forming the transfer matrix of a lattice theory with gauge and matter fields. To sum up these operators include the vertex, plaquette and link operators along with the single qudit operators acting in the gauge sector given by Eq.(\ref{algZ}) and Eq.(\ref{algX}) and in the matter sector given by Eq.(\ref{algQ}). We now proceed to study the algebra between these operators. 

\subsection{Algebra of the Operators}

The remaining basic operators in the theory are given by $L_i^g$, $R_i^g$, $T_i^g$ and $Q_v^{i}$. Clearly $T_i^g$ and $Q_v^{i}$ are projectors and $Q_v^{i}$ commutes with the remaining operators as they act on different sectors. We have the following relations
\bea L_i^{g_1}L_i^{g_2} & = & L_i^{g_1g_2} \\
R_i^{g_1} R_i^{g_2} & = & R_i^{g_2g_1} \\ 
T_i^{g_2}L_i^{g_1} & = & L_i^{g_1}T_i^{g_1^{-1}g_2} \\ 
T_i^{g_2}R_i^{g_1} & = & R_i^{g_1}T_i^{g_2g_1^{-1}}.\eea
Using these relations it is easy to see that the vertex and plaquette operators given by Eq.(\ref{algAv}) and Eq.(\ref{algBp}) commute. This computation is exactly similar to the computation used to prove the commutativity of these two operators in the case of the quantum double models. This is true because the plaquette operators in our case of theories with gauge and matter fields is left unchanged with respect to the case with pure gauge fields.

The link operators $C_l$ trivially commute with the plaquette operators as both are diagonal. This can be seen from their respective definitions given by Eq.(\ref{algC}) and Eq.(\ref{algBp}). The only thing that we have to prove is the commutation between the vertex operator and the link operator in their common region of support. 

Consider the action of the operators $A_v$ and $C_l$ on the gauge and matter degrees of freedom. The proof then goes as follows for their action on the common support, we distinguish two cases as the vertex operator action depends on the orientation of the lattice, the first case being the one when the vertex operator acts on the vertex at the left of the edge $l$, this is
\begin{align}\label{comm1}
 A_{v_1}^{g} C_l \Ket{\dots,\chi_{v_1},\phi_l,\chi_{v_2},\dots}&=\langle\mu(\phi_l)\chi_{v_1},\chi_{v_2}\rangle_{G}A_{v_1}^{g}\Ket{\dots,\chi_{v_1},\phi_l,\chi_{v_2},\dots}\nonumber\\
 &=\langle\mu(\phi_l)\chi_{v_1},\chi_{v_2}\rangle_{G}\Ket{\dots,\mu(\phi_g)\chi_{v_1},\phi_l\phi_{g^{-1}},\chi_{v_2},\dots},
\end{align}
whereas
\begin{align}\label{comm2}
 C_l A_{v_1}^{g} \Ket{\dots,\chi_{v_1},\phi_l,\chi_{v_2},\dots}&=C_l \Ket{\dots,\mu(\phi_g)\chi_{v_1},\phi_l\phi_{g^{-1}},\chi_{v_2},\dots}\nonumber\\
 &=\langle\mu(\phi_l\phi_{g^{-1}})\mu(\phi_g)\chi_{v_1},\chi_{v_2}\rangle_{G}\Ket{\dots,\mu(\phi_g)\chi_{v_1},\phi_l\phi_{g^{-1}},\chi_{v_2},\dots},
\end{align}
since the module map is a group homomorphism we note that:
\begin{align*}
 \mu(\phi_l\phi_{g^{-1}})\mu(\phi_g)&=\mu(\phi_l)\mu(\phi_{g^{-1}})\mu(\phi_g)\\
 &=\mu(\phi_l)\mu(\phi_{g^{-1}}\phi_g)\\
 &=\mu(\phi_l),
\end{align*}
thus, the coefficient on the right hand side of eq.(\ref{comm1}) now is the same as the one in eq.(\ref{comm2}), hence $[A_{v_1}^{g},C_l]=0$. Let us now consider the case when the vertex operator is acting at the right of the edge $l$, i.e.
\begin{align}\label{comm3}
 A_{v_2}^{g} C_l \Ket{\dots,\chi_{v_1},\phi_l,\chi_{v_2},\dots}&=\langle\mu(\phi_l)\chi_{v_1},\chi_{v_2}\rangle_{G} A_{v_2}^{g} \Ket{\dots,\chi_{v_1},\phi_l,\chi_{v_2},\dots}\nonumber\\
 &=\langle\mu(\phi_l)\chi_{v_1},\chi_{v_2}\rangle_{G}\Ket{\dots,\chi_{v_1},\phi_g\phi_l,\mu(\phi_g)\chi_{v_2},\dots},
\end{align}
while
\begin{align}\label{comm4}
 C_l A_{v_2}^{g} \Ket{\dots,\chi_{v_1},\phi_l,\chi_{v_2},\dots}&=C_l \Ket{\dots,\chi_{v_1},\phi_g\phi_l,\mu(\phi_g)\chi_{v_2},\dots}\nonumber\\
 &=\langle\mu(\phi_g\phi_l)\chi_{v_1},\mu(\phi_g)\chi_{v_2}\rangle_{G}\Ket{\dots,\chi_{v_1},\phi_g\phi_l,\mu(\phi_g)\chi_{v_2},\dots},
\end{align}
for the commutation to hold we require:
\begin{equation}
 \langle\mu(\phi_g)\chi_{\alpha},\chi_{\beta}\rangle_{G}=\langle\chi_{\alpha},\mu(\phi_g)^{\dagger}\chi_{\beta}\rangle_{G},
\end{equation}
this means the representation map $\mu$ is unitary, equivalently
\begin{equation}
 \mu(\phi_g)^{\dagger}=\mu(\phi_{g^{-1}}).
\end{equation}
therefore the right hand side of eq.(\ref{comm4}) reduces to that of eq.(\ref{comm3}). Thus, $[A_{v_2}^g,C_l]=0$.
So, we just showed that $[A_v^g,C_l]=0$ for any $\phi_g \in \mathbb{C}(G) $ and for any vertex $v$ of the lattice, thus it follows that $[A_v,C_l]=0$. 

Moreover we only consider $\mu$'s which permute the basis elements of the matter module. Under this condition $\langle\mu(\phi_l).\chi_{v_1}|\chi_{v_2}\rangle$ is real which makes the link operator $C_l$ hermitian. This also ensures that the vertex operator is hermitian as the gauge part of the vertex operator is the same as the one in the QDM of Kitaev. The plaquette operator is the same as those appearing in the QDM of Kitaev. Thus all the operators in the Hamiltonian are hermitian making the evolution unitary.

\subsection{The Transfer Matrix}

We are now in a position to write down the full transfer matrix of a lattice theory with gauge and matter fields as a product of these local operators.  Thus we have
\begin{eqnarray}
U\left(z_\textrm{S},z_\textrm{T},z_\textrm{S}^*,z_\textrm{T}^*,m_V,G_S,G_T \right) &=& \prod_p B_p(z_S) \prod_l C_l(G_S)  \prod_l\left(T_l(z_S^*)D_l L_l(z_T) \right) \nn \\
&~&\prod_v\tilde{V}_v(G_T) Q_v(m_v) A_v(z_T^*)\nn
\end{eqnarray}
%
%
%
The other operators namely $B_p(z_S)$, $T_l(z_S^*)$, $Q_v(m_v)$, $L_l(z_T)$, $A_v(z_T^*)$ are given by Eq.(\ref{algBp}), Eq.(\ref{algZ}), Eq.(\ref{algQ}), Eq.(\ref{algX}), Eq.(\ref{algAv}) respectively. Through these operators the transfer matrix $U$ obtains its dependence on $z_S$, $z_S^*$, $m_v$, $z_T$ and $z_T^*$ respectively. 

The operator $\tilde{C}_l$ is defined as a linear combination of orthogonal projectors. We write it as 
\beq \tilde{C}_l = \gamma C_l + \gamma_\perp C_l^\perp \eeq
where $C_l$ is given by Eq.(\ref{algC}) and $C_l^\perp$ is the operator orthogonal to $C_l$. The form of the orthogonal operator $C_l^\perp$ depends on the module $H_n$ considered. In general there are several operators orthogonal to the operator $C_l$.  For the modules with symmetry group as $\mathbb{Z}_n$ we can write down a general form for these orthogonal operators. If we denote the inner product in the module $H_n$ as $\langle\chi_i|\chi_j\rangle$ then we can write down the operator $C_l$ as
\beq C_l|\chi_{v_1}, \phi_l, \chi_{v_2}\rangle = \langle\mu(\phi_l).\chi_{v_1}|\chi_{v_2}\rangle |\chi_{v_1}, \phi_l, \chi_{v_2}\rangle. \eeq  
We can now write $n-1$ other orthogonal operators as 
\beq \label{algCorth}C_l^{\perp}|\chi_{v_1}, \phi_l, \chi_{v_2}\rangle = \langle\mu(\phi_l).\chi_{v_1}|X^i|\chi_{v_2}\rangle |\chi_{v_1}, \phi_l, \chi_{v_2}\rangle;~ i\in \{1,\cdots, n-1\}\eeq
where $X$ is the shift operator generating $\mathbb{Z}_n$ given by 
\beq X^k|\chi_i\rangle = |\chi_{i+k}\rangle.\eeq
The transfer matrix gets its dependence on $G$ through the link operator $C_l$. The inner product given by $G$ in turn depends on the matrix $R$ as shown by it's tensor network representation in figure (\ref{pc8}). 

Thus for each $i$ in Eq.(\ref{algCorth}) we choose a different $R$ matrix in the inner product $G$. For a given $i$ in Eq.(\ref{algCorth}) $R$ is given by $(X^i)^{\frac{1}{2}}$. This matrix $R$ can be thought of as a basis transformation in the module $H_n$. 

Having written down the general expression for the transfer matrix $U$ of a lattice theory with gauge and matter fields we will now consider specific examples which illustrate the formalism developed so far.

\section{Examples}

We will consider two examples of Hamiltonians having long-ranged entangled ground states derived from the construction we have illustrated. The first example is the simplest Hamiltonian with the gauge group being $\mathbb{Z}_2$ and the vector space carrying its representation being the two dimensional vector space $H_2$. We will denote this model as $H_2/\mathbb{Z}_2$. The second example is a slight modification with the vector space $H_2$ replaced by $H_3$, a three dimensional vector space carrying the representation of the gauge group $\mathbb{Z}_2$. We denote this model as $H_3/\mathbb{Z}_2$. We consider each of them separately in what follows. 

The ground state degeneracies in each of these cases is a topological invariant and can be computed numerically. There are no winding operators as in the toric code for the $H_2/\mathbb{Z}_2$ case but we do have a winding operator for the $H_3/\mathbb{Z}_2$ case. We will write down the ground states in both the cases and give their tensor network representations.

Both these examples have confined excitations apart from one deconfined excitation. We will briefly look at the vertex and gauge excitations in the $H_2/\mathbb{Z}_2$ case. The vertex excitations of $H_3/\mathbb{Z}_2$ were studied in detail in~\cite{pp2}. Here we will just look at it's deconfined excitation. We then present a third example, the $H_2/\mathbb{Z}_4$ model where we use an action of $\mathbb{Z}_4$ on $H_2$ such that one of the fluxes become deconfined even in the presence of the link operator $C_l$. 

The effect of confinement is due to the link operator $C_l$. A systematic way to deconfine all the fluxes is by working with Hamiltonians which do not have the $C_l$ operators but just the vertex operator $A_v$ and the plaquette operator $B_p$. We then see that we have all the deconfined excitations of the quantum double models along with the vertex excitations due to the presence of the matter fields. These models will comprise the fourth set of examples. This is the simplest and most straightforward way to deconfine all the fluxes in contrast with the $H_2/\mathbb{Z}_4$ case where a clever choice of input data and structure constant helped us deconfine the fluxes.  

Finally we will show how to recover the quantum double Hamiltonians with only gauge fields by ``switching off'' the matter fields. These comprise the fifth set of examples. All the examples are constructed on the torus.

Before we write down the models we make a few comments about the ground state degeneracy of these models. The models we will construct are obtained from the transfer matrix which has the following form
\beq U = \prod_p B_p\prod_lC_l\prod_vA_v \eeq which makes this operator a projector. Using the arguments of~\cite{pp1} we can show that the trace of this operator gives the ground state degeneracy. With a little more work it is possible to show that this is also a topological invariant. This requires the introduction of Pachner moves in the presence of matter fields. We reserve this discussion for another paper. 

The trace of the transfer matrix can be computed numerically\footnote{We thank Kazuo Teramoto for these computations.} and these results are used for the two examples to be discussed below.

\subsection{$H_2/\mathbb{Z}_2$:}

The basis elements of the group algebra of $\mathbb{Z}_2$ are denoted by the elements of the set $\{\phi_1, \phi_{-1}\}$ and the basis elements of the module $H_2$ are denoted by $\{\chi_1, \chi_{-1}\}$. The element $\phi_1$ acts as the identity element on the module $H_2$. The element $\phi_{-1}$ on the other hand flips between the two elements of $H_2$. That is the representation map $\mu(\phi_{-1})$ is given by
\bea \mu(\phi_{-1}).\chi_1 & = & \chi_{-1} \\
 \mu(\phi_{-1}).\chi_{-1} & = & \chi_1.\eea
As a matrix $\mu(\phi_{-1})$ is the $\sigma^x$ Pauli matrix and $\mu(\phi_1)$ is the two by two identity matrix. Henceforth we will use the Pauli matrices in place of the representation matrices. 

The two orthogonal vertex operators are given by
\bea \label{av1}A_v^1 & = & \frac{\mathbb{I}\otimes \mathbb{I}\otimes \mathbb{I}\otimes \mathbb{I}\otimes \mathbb{I} + \sigma^x_v\otimes \sigma^x_{l_1}\otimes \sigma^x_{l_2}\otimes \sigma^x_{l_3}\otimes \sigma^x_{l_4}}{2} \\
A_v^{-1} & = & \frac{\mathbb{I}\otimes \mathbb{I}\otimes \mathbb{I}\otimes \mathbb{I}\otimes \mathbb{I} - \sigma^x_v\otimes \sigma^x_{l_1}\otimes \sigma^x_{l_2}\otimes \sigma^x_{l_3}\otimes \sigma^x_{l_4}}{2}. \eea

The plaquette operators are the same as in the toric code case and are given by
\bea \label{bp1}B_p^1 & = & \frac{\mathbb{I}\otimes \mathbb{I}\otimes \mathbb{I}\otimes \mathbb{I} + \sigma^z_{l_1}\otimes \sigma^z_{l_2}\otimes \sigma^z_{l_3}\otimes \sigma^z_{l_4}}{2} \\ 
B_p^{-1} & = & \frac{\mathbb{I}\otimes \mathbb{I}\otimes \mathbb{I}\otimes \mathbb{I} - \sigma^z_{l_1}\otimes \sigma^z_{l_2}\otimes \sigma^z_{l_3}\otimes \sigma^z_{l_4}}{2}.\eea
As can be seen from these expressions the plaquette operator acts trivially on the matter sector. 

There are two orthogonal link operators in this case. Their action on the basis states of the system are given by
\bea\label{cl1} C_l^1|\chi_{v_1}, \phi_l, \chi_{v_2}\rangle & = & \langle\mu(\phi_l).\chi_{v_1}|\chi_{v_2}\rangle |\chi_{v_1}, \phi_l, \chi_{v_2}\rangle \\
\label{cl2} C_l^{-1}|\chi_{v_1}, \phi_l, \chi_{v_2}\rangle & = & \langle\mu(\phi_l).\chi_{v_1}|\sigma^x|\chi_{v_2}\rangle |\chi_{v_1}, \phi_l, \chi_{v_2}\rangle \eea
We now obtain the matrix representations of these two orthogonal operators. This is done by summing the projectors onto all the configurations that satisfy the constraint given by the inner product. 
The set of configurations $\chi_{v_1}$, $\phi_l$ and $\chi_{v_2}$ which make the inner product for $C_l^1$ non-zero is shown in table (\ref{sometable}). 
\begin{table}
\begin{center}
\begin{tabular}{|c|c|c|}
\hline
$v_1$ & $l$ & $v_2$ \\
\hline
1 & 1 & 1 \\
-1 & -1 & 1 \\
-1 & 1 & -1 \\
1 & -1 & -1 \\
\hline
\end{tabular}
\end{center}
\caption{Configurations of $\chi_{v_1}$, $\phi_l$ and $\chi_{v_2}$ which make the inner product in Eq.(\ref{cl1}) non-zero. } \label{sometable}
\end{table}
The projector to these configurations is written as the following sum
\bea C_l^1 & = & \left(\frac{1+\sigma^z_{v_1}}{2}\right)\otimes\left(\frac{1+\sigma^z_l}{2}\right)\otimes\left(\frac{1+\sigma^z_{v_2}}{2}\right) \nn \\
& + & \left(\frac{1-\sigma^z_{v_1}}{2}\right)\otimes\left(\frac{1-\sigma^z_l}{2}\right)\otimes\left(\frac{1+\sigma^z_{v_2}}{2}\right) \nn \\
& + & \left(\frac{1-\sigma^z_{v_1}}{2}\right)\otimes\left(\frac{1+\sigma^z_l}{2}\right)\otimes\left(\frac{1-\sigma^z_{v_2}}{2}\right) \nn \\
& + & \left(\frac{1+\sigma^z_{v_1}}{2}\right)\otimes\left(\frac{1-\sigma^z_l}{2}\right)\otimes\left(\frac{1-\sigma^z_{v_2}}{2}\right). \eea
This reduces to 
\beq C_l^1 = \frac{1\otimes 1 \otimes 1 + \sigma^z_{v_1}\otimes\sigma^z_l\otimes\sigma^z_{v_2}}{2}.\eeq

In a similar way we can compute $C_l^{-1}$ which is the orthogonal projector to $C_l^1$. The set of configurations $\chi_{v_1}$, $\phi_l$ and $\chi_{v_2}$ which make the inner product for $C_l^{-1}$ non-zero is shown in table (\ref{sometable1}). 

\begin{table}
\begin{center}
\begin{tabular}{|c|c|c|}
\hline
$v_1$ & $l$ & $v_2$ \\
\hline
-1 & 1 & 1 \\
1 & -1 & 1 \\
1 & 1 & -1 \\
-1 & -1 & -1 \\
\hline
\end{tabular}
\end{center}
\caption{Configurations of $\chi_{v_1}$, $\phi_l$ and $\chi_{v_2}$ which make the inner product in Eq.(\ref{cl2}) non-zero. } \label{sometable1}
\end{table}

The projector to these configurations is written as the following sum
\bea C_l^{-1} & = & \left(\frac{1+\sigma^z_{v_1}}{2}\right)\otimes\left(\frac{1+\sigma^z_l}{2}\right)\otimes\left(\frac{1-\sigma^z_{v_2}}{2}\right) \nn \\
& + & \left(\frac{1-\sigma^z_{v_1}}{2}\right)\otimes\left(\frac{1-\sigma^z_l}{2}\right)\otimes\left(\frac{1-\sigma^z_{v_2}}{2}\right) \nn \\
& + & \left(\frac{1-\sigma^z_{v_1}}{2}\right)\otimes\left(\frac{1+\sigma^z_l}{2}\right)\otimes\left(\frac{1+\sigma^z_{v_2}}{2}\right) \nn \\
& + & \left(\frac{1+\sigma^z_{v_1}}{2}\right)\otimes\left(\frac{1-\sigma^z_l}{2}\right)\otimes\left(\frac{1+\sigma^z_{v_2}}{2}\right). \eea
This reduces to 
\beq C_l^{-1} = \frac{1\otimes 1 \otimes 1 - \sigma^z_{v_1}\otimes\sigma^z_l\otimes\sigma^z_{v_2}}{2}.\eeq

Using the matrix representations for all the operators it is easy to see that they indeed commute with each other and they are all projectors. The model posses a global symmetry given by the operator $\prod_v\sigma^x_v$ where the product runs over all the vertices on the lattice. This can also be thought of as the symmetry group of the module $H_2$. This comparison will be crucial later on.  

The operator $X_l$ is given by 
\beq X_l = a_1\left(\frac{1 + \sigma^x_l}{2}\right) + a_{-1} \left(\frac{1 - \sigma^x_l}{2}\right). \eeq

The operator $Z_l$ is given by 
\beq Z_l = b_1\left(\frac{1 + \sigma^z_l}{2}\right) + b_{-1} \left(\frac{1 - \sigma^z_l}{2}\right). \eeq

The operator $Q_v$ is given by 
\beq Q_v = c_1\left(\frac{1 + \sigma^z_v}{2}\right) + c_{-1} \left(\frac{1 - \sigma^z_v}{2}\right). \eeq

The connector operator $V_v$ is given by 
\beq V_v = d_1\left(\frac{1 + \sigma^x_v}{2}\right) + d_{-1} \left(\frac{1 - \sigma^x_v}{2}\right). \eeq

The other connector operator $L_l$ is proportional to identity~\cite{pp1}. Thus we have the full transfer matrix for the $H_2/\mathbb{Z}_2$ case. We can now use the definition of the transfer matrix $$ U\left(z_S, z_T, z_S^*, z_T^*, m_v, G_S, G_T\right) = e^{-H}$$ to find several Hamiltonians having $H_2$ and $\mathbb{Z}_2$ degrees of freedom. One such example which is exactly solvable is given by the Hamiltonian
\beq\label{h2} H = -\sum_v A_v - \sum_p B_p - \sum_l C_l.\eeq
We now study the ground states of this Hamiltonian. The Hamiltonian in Eq.(\ref{h2}) is exactly solvable as the operators making up the Hamiltonian are commuting projectors. Since the eigenvalues of these projectors are 0 or 1 the entire spectrum of the Hamiltonian is known. The condition for the ground states in particular is given by 
\beq\label{cond2} A_v|gr\rangle = B_p|gr\rangle = C_l|gr\rangle = |gr\rangle;~ \forall v, p,l.\eeq
This gives the lowest energy eigenvalue as $-(N_v+N_p+N_l)$, where $N_v$, $N_p$ and $N_l$ are the number of vertices, plaquettes and links respectively. Numerical computations of the trace of the transfer matrix in this case shows that the ground state degeneracy is one in this case. 

\begin{figure}[h!]
\begin{center}
		\includegraphics[scale=1]{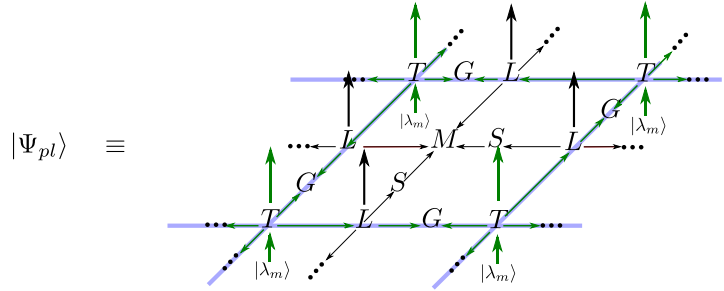}
	\caption{The tensor network representation of the state $|\psi_{pl}\rangle$ given in Eq.(\ref{gpl2}).}
\label{psipl}
\end{center}
\end{figure}

One such state satisfying the conditions of Eq.(\ref{cond2}) is given by 
\beq\label{gpl2} |\psi_{pl}\rangle = \prod_pB_p\prod_lC_l \otimes_l|\lambda_g\rangle\otimes_v|\lambda_m\rangle \eeq
where $\lambda_g = \phi_1 + \phi_{-1}$ and $\lambda_m = \chi_1 + \chi_{-1}$. The representation of this state as a tensor network is shown in figure (\ref{psipl}). 

\begin{figure}[h!]
\begin{center}
		\includegraphics[scale=1]{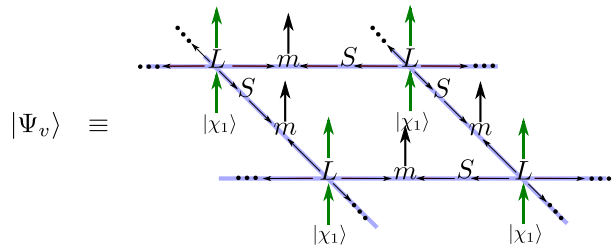}
	\caption{The tensor network representation of $|\psi_v\rangle$ given in Eq.(\ref{gv}).}
\label{psiv}
\end{center}
\end{figure}

We can write another ground state for this model as
\beq\label{gv} |\psi_v\rangle = \prod_v A_v\otimes_l |\phi_1\rangle\otimes_v|\chi_1\rangle. \eeq 
The tensor network representation of this state is shown in figure (\ref{psiv}).

The two ground states $|\psi_{pl}\rangle$ and $|\psi_v\rangle$ are similar to the two ground states in the case of the toric code as discussed in~\cite{pp1}. They can be thought as being written in the basis of $\sigma^x$ and $\sigma^z$ respectively. From our numerical considerations we presume that these two states are the same. Another way to see this is that there is only one equivalence class under the gauge action, that is the representation vector space has a single orbit under $\mathbb{Z}_2$ action given by $H_2$ itself. Another reason why we expect these states to be the same is that there are no winding operators in this case and we know from our argument in~\cite{pp1} that using the presence of winding operators we can relate the two basis sets of states.   

In this model we have a third possibility for a ground state, formed out of a mixture of states from the $\sigma^z$ basis and the dual basis, given by 
\beq \label{gvl} |\psi_{vl}\rangle = \prod_vA_v\prod_lC_l\otimes_l |\phi_1\rangle\otimes_v|\lambda_m\rangle. \eeq
As the ground state degeneracy is one and there are no winding operators we again expect this state to be the same as $|\psi_v\rangle$ and $|\psi_{pl}\rangle$. 

\begin{figure}[h!]
 \centering
 \includegraphics[scale=1]{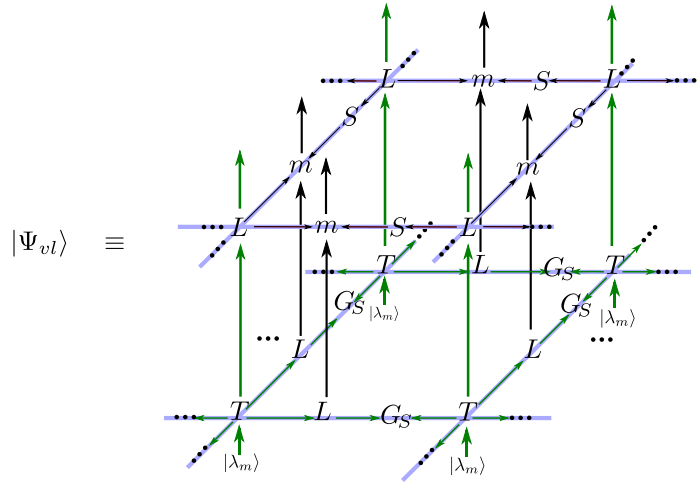}
 \caption{The tensor network representation of the state $\ket{\psi_{vl}}$ of Eq.~\ref{gvl}.}\label{psivl}
\end{figure}

\begin{figure}[h!]
 \centering
 \includegraphics[scale=1]{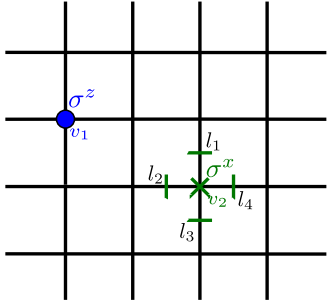}
 \caption{The vertex excitations in the matter sector of the $H_2/\mathbb{Z}_2$ model.}\label{vertex}
\end{figure}

We now look at the excitations in this model. There are both gauge and vertex excitations in this model. The vertex excitations are obtained by applying $\sigma^z_v$ and $\sigma^x_v$ vertex $v$. The former commutes with the link operator and the plaquette operator but does not commute with the vertex operator $A_v$ creating a single vertex excitation and the latter commutes with the vertex and plaquette operator but does not commute with the four link operators adjoining a vertex $v$, thus creating four link excitations. Clearly these excitations are not anyonic like in the toric code case. The link excitations in fact depend on the valency of the vertices in the lattice. These excitations are shown in figure (\ref{vertex}).

\begin{figure}[h!]
 \centering
 \includegraphics[scale=1]{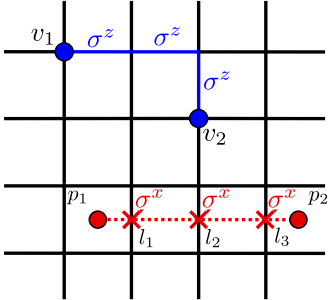}
 \caption{The gauge excitations in the $H_2/\mathbb{Z}_2$ model showing the deconfined charges and the confined fluxes.}\label{gaugeex}
\end{figure}

The gauge excitations are obtained by applying $\sigma^z_l$ and $\sigma^x_l$ on the links of the lattice. The former commutes with the plaquette and link operators and creates charge excitations just as in the toric code case. They are deconfined in the sense that applying a string of $\sigma^z$ operators on the direct lattice creates a pair of charges at the end points of the string irrespective of the size of the string. This feature is not present for the flux excitations as every time we apply a $\sigma^x_l$ on a link we also create a link excitation on the link $l$. Thus if we separate a pair of fluxes by applying a string of $\sigma^x_l$ operators on the dual lattice we end up creating a series of link excitations along the way. Thus the fluxes in this model are confined by the string tension provided by the link operator $C_l$. This is shown in figure (\ref{gaugeex}).    

\begin{figure}[h!]
 \centering
 \includegraphics[scale=1]{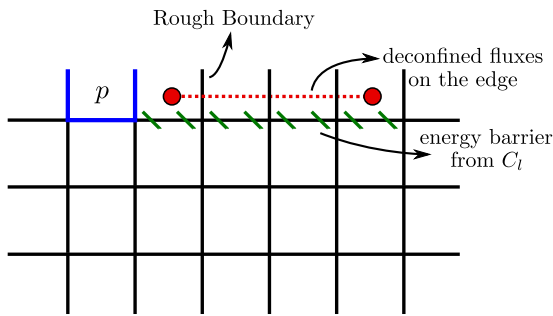}
 \caption{Deconfined fluxes localized at the rough boundary. }\label{rough}
\end{figure}

We now consider placing this model on a two dimensional lattice with a rough boundary. This changes the Hamiltonian at the boundary in such a way that they model is still exactly solvable. The bulk Hamiltonian remains the same and is given by Eq.~\ref{h2}. The boundary Hamiltonian is given by 
\beq H_{\textrm{rough boundary}} = -\sum_{p\in \partial\mathcal{M}} \tilde{B}_p\eeq
with $\tilde{B}_p$ being the plaquette operators for the three sided plaquettes given by 
\beq \tilde{B}_p = \frac{1\otimes 1\otimes 1 + \sigma^z\otimes\sigma^z\otimes\sigma^z}{2}.\eeq
These operators are shown in the lattice in figure (\ref{rough}). It is clear that there cannot be any link operators on the rough boundary and hence the fluxes are now deconfined on the edge. If we try to move them into the bulk we have to overcome the energy barrier created by the link operator. Thus they are localized on the edges and are gapped.

\begin{figure}[h!]
 \centering
 \includegraphics[scale=1]{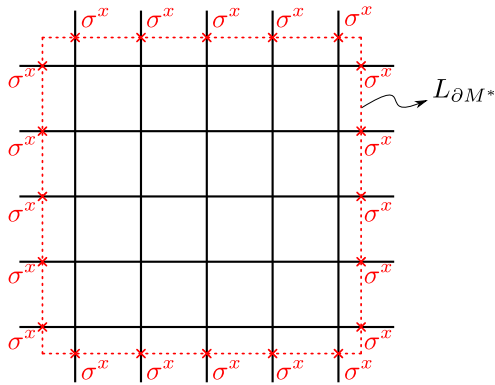}
 \caption{The loop operator acts on the non-degenerate ground state to give a new one. }\label{loop}
\end{figure}

 The existence of deconfined fluxes at the rough edge implies that the ground state degeneracy (GSD) of the system now increases. This is due to the loop operator 
 \beq L^x = \prod_{i\in L_{\partial M^*}} \sigma^x_i \eeq
 shown in figure (\ref{loop}).
Thus the $H_2/\mathbb{Z}_2$ system when placed on a manifold with rough boundary develops a ground state degeneracy in contrast to the situation when it is placed on a closed manifold. 

This feature is similar to the ones occurring in ``confined Walker-Wang'' models~\cite{ww} as elaborated in~\cite{burst}. These models are exactly solvable 3D lattice models with a confined bulk but with deconfined anyons on the surface. Such surface states were also seen to be outside the group cohomology classification of SPT phases~\cite{wenSPT} in~\cite{ash2, ash3}. The $H_2/\mathbb{Z}_2$ presented here is a 2D model with this feature, the difference being that, only the fluxes are confined in the bulk and they get deconfined on rough boundaries. We believe these also to be outside the group cohomology classification scheme of SPT phases as the global symmetry does not act with an obstruction on these states~\cite{nayak3}. Though we have seen this in this particular example it is easy to show that this is a property of all the systems we obtain from this formalism. In particular they are also true for non-Abelian groups. We will discuss this aspect of this model in another paper.

\subsection{$H_3/\mathbb{Z}_2$:}

The module $H_3$ is spanned by the basis elements $\{\chi_0, \chi_1, \chi_2\}$. The identity element $\phi_1$ of the gauge group $\mathbb{Z}_2$ acts as identity on $H_3$ and is thus the three by three identity matrix. The other element $\phi_{-1}$ flips $\chi_0$ and $\chi_1$ and leaves $\chi_2$ invariant. This makes the matrix representation of $\mu(\phi_{-1})$ as
\beq\label{mu32} \mu_v(\phi_{-1}) = \left(\begin{array}{ccc} 0 & 1 & 0 \\ 1 & 0 & 0 \\ 0 & 0 & 1\end{array}\right).\eeq 
Note that we can define other permutation actions of the gauge group $\mathbb{Z}_2$ on $H_3$ where $\mu(\phi_{-1})$ leaves either $\chi_0$ or $\chi_1$ invariant instead of $\chi_2$. These actions are unitarily equivalent to the one defined above which leaves $\chi_2$ invariant. We will work with the one given in Eq.~\ref{mu32}. 

The vertex operator in this model is given by
\beq\label{star1} A_v = \frac{1\otimes1\otimes1\otimes1\otimes1 + \mu_v(\phi_{-1})\otimes\sigma^x_{i_1}\otimes\sigma^x_{i_2}\otimes\sigma^x_{i_3}\otimes\sigma^x_{i_4}}{2} \eeq 
where $\sigma^x$ is the Pauli matrix. The orthogonal operators become important in finding the excitations and hence we will also write them down. The operator orthogonal to Eq.(\ref{star1}) is given by 
\beq\label{star2} A_v^\perp = \frac{1\otimes1\otimes1\otimes1\otimes1 - \mu_v(\phi_{-1})\otimes\sigma^x_{i_1}\otimes\sigma^x_{i_2}\otimes\sigma^x_{i_3}\otimes\sigma^x_{i_4}}{2}. \eeq
 
 The link operator $C_l$ is given by 
\begin{eqnarray}\label{link32} C_l & = & \left[\frac{1 + Z_{v_1} + Z_{v_1}^2}{3}\right]\otimes\left[\frac{1+\sigma^z_l}{2}\right]\otimes\left[\frac{1 + Z_{v_2} + Z_{v_2}^2}{3}\right] \nn \\
                                  & + & \left[\frac{1 + \omega^2Z_{v_1} + \omega Z_{v_1}^2}{3}\right]\otimes\left[\frac{1-\sigma^z_l}{2}\right]\otimes\left[\frac{1 + Z_{v_2} + Z_{v_2}^2}{3}\right] \nn \\
                                  & + & \left[\frac{1 + Z_{v_1} + Z_{v_1}^2}{3}\right]\otimes\left[\frac{1-\sigma^z_l}{2}\right]\otimes\left[\frac{1 + \omega^2Z_{v_2} + \omega Z_{v_2}^2}{3}\right] \nn \\ 
                                  & + & \left[\frac{1 + \omega^2Z_{v_1} + \omega Z_{v_1}^2}{3}\right]\otimes\left[\frac{1+\sigma^z_l}{2}\right]\otimes\left[\frac{1 +\omega^2 Z_{v_2} +\omega Z_{v_2}^2}{3}\right] \nn \\
                                  & + & \left[\frac{1 + \omega Z_{v_1} + \omega^2Z_{v_1}^2}{3}\right]\otimes1\otimes\left[\frac{1 + \omega Z_{v_2} + \omega^2Z_{v_2}^2}{3}\right] \end{eqnarray} 
where $Z$ is a generator of $\mathbb{Z}_3$ given by 
\beq\label{z4} Z = \left(\begin{array}{ccc} 1 & 0 & 0 \\ 0 & \omega & 0 \\ 0 & 0 & \omega^2 \end{array}\right) \eeq
and $\omega = e^{i\frac{2\pi}{3}}$.  

The plaquette operator is the same as the previous example. 

Once again we can write down exactly solvable Hamiltonians of the form  
\beq\label{h2} H = -\sum_v A_v - \sum_p B_p - \sum_l C_l\eeq
with $A_v$ given by Eq.(\ref{star1}) and $C_l$ given by Eq.(\ref{link32}) respectively. 

The ground states of this model satisfies the condition $A_v|gr\rangle = B_p|gr\rangle = C_l|gr\rangle=|gr\rangle~\forall v,p,l$. This is the same as in the previous example and is due to the fact that the Hamiltonian is made up of commuting projectors which leaves the spectrum of the model unchanged with respect to the previous example. Numerical computation of the trace of the transfer matrix gives us 5 as the ground state degeneracy for this model. We can write down the 5 ground states of this model as 
\bea \label{grH3z2} |\psi_v, 0\rangle  & = & \prod_vA_v \otimes_l|\phi_1\rangle\otimes_v|\chi_0\rangle \\ 
|\psi_v, 2\rangle  & = & \prod_vA_v \otimes_l|\phi_1\rangle\otimes_v|\chi_2\rangle \\
|\psi_v, 2, C_1^*\rangle  & = & \prod_vA_v \otimes_{l\notin C_1^*}|\phi_1\rangle\otimes_{l\in C_1^*}|\phi_{-1}\rangle\otimes_v|\chi_2\rangle \\
|\psi_v, 2, C_2^*\rangle  & = & \prod_vA_v \otimes_{l\notin C_2^*}|\phi_1\rangle\otimes_{l\in C_2^*}|\phi_{-1}\rangle\otimes_v|\chi_2\rangle \\
|\psi_v, 2, C_1^*,C_2^*\rangle  & = & \prod_vA_v \otimes_{l\notin C_1^*,C_2^*}|\phi_1\rangle\otimes_{l\in C_1^*,C_2^*}|\phi_{-1}\rangle\otimes_v|\chi_2\rangle \eea
where $C_1^*$ and $C_2^*$ are non-contractible loops along the dual lattice around the two independent directions of the torus. It is easy to see that these states are linearly independent. Note that the first two states in this list have the gauge transformations acting in equivalence classes of the gauge group acting on the vector space carrying its representation, that is $\mathbb{Z}_2$ acting on $H_3$. It can be seen from the gauge action that the equivalence classes in this case are given by $\{\chi_0, \chi_1\}$ and $\{\chi_2\}$. We can also see that within an equivalence class the state resembles that of the quantum double of the gauge group that is of the toric code in this case.  

We can write down another set of 5 ground states in the dual basis which are linear combinations of the above states. These are given by
\bea \label{grH3z2dual} |\psi_p\rangle & = &\prod_p B_p\prod_lC_l \otimes_l|\phi_1 + \phi_{-1}\rangle\otimes_v|\chi_0+\chi_1+ \chi_2\rangle \\
|\psi_, C_1\rangle & = & \prod_p B_p\prod_lC_l \prod_{j\in C_1}\sigma^z_j\otimes_l|\phi_1 + \phi_{-1}\rangle\otimes_v|\chi_0+\chi_1+ \chi_2\rangle \\
|\psi_, C_2\rangle & = &\prod_p B_p\prod_lC_l \prod_{j\in C_2}\sigma^z_j\otimes_l|\phi_1 + \phi_{-1}\rangle\otimes_v|\chi_0+\chi_1+ \chi_2\rangle \\
|\psi_, C_1,C_2\rangle & = &\prod_p B_p\prod_lC_l \prod_{j\in C_1, C_2}\sigma^z_j\otimes_l|\phi_1 + \phi_{-1}\rangle\otimes_v|\chi_0+\chi_1+ \chi_2\rangle \\
|\psi_{vl}\rangle & = & \prod_v A_v\prod_l C_l \otimes_l|\phi_1\rangle \otimes_v|\chi_0+\chi_1+\chi_2\rangle \eea
where $C_1$ and $C_2$ are non-contractible curves along the direct lattice around the two independent directions of the torus. For example we can see that $|\psi_p\rangle = |\psi_v, 0\rangle +  |\psi_v, 2\rangle + |\psi_v, 2, C_1^*\rangle + |\psi_v, 2, C_2^*\rangle + |\psi_v, 2, C_1^*,C_2^*\rangle$. The other states obtained by applying the winding operator are similar combinations of the vertex states. 

Placing this system on a manifold with boundary gives new edge states, similar to the $H_2/\mathbb{Z}_2$ case. The confined flux, in the $\{\chi_0\}$ equivalence class on the closed manifold, gets deconfined on a rough boundary. Apart from this deconfined flux we also have the deconfined flux in the $\{\chi_3\}$ equivalence class. However the ground state degeneracy decreases from 5 on the closed manifold to 3 on a manifold with a rough boundary.  

\subsection{$H_2/\mathbb{Z}_4$:}

We now consider another model with $\mathbb{Z}_4$ gauge fields acting on the two dimensional vector space $H_2$. The basis elements of $\mathbb{Z}_4$ are denoted by $\{\phi_0, \phi_1, \phi_2, \phi_3\}$ and those of $H_2$ are $\{\chi_1, \chi_{-1}\}$. The gauge action is given by the following two by two matrices, $$\mu(\phi_0)=\mu(\phi_2) = \left(\begin{array}{cc} 1 & 0 \\ 0 & 1 \end{array}\right)$$ and 
  $$\mu(\phi_1)=\mu(\phi_3) = \left(\begin{array}{cc} 0 & 1 \\ 1 & 0 \end{array}\right).$$
This model has one extra deconfined flux in addition to the three deconfined charges. This can be seen due to the trivial action of the gauge group element $\phi_2$ on $H_2$. 

In terms of operators it is the statement that the operator $X_l^2$, where $X$ is the shift operator given by $$ X = \left(\begin{array}{cccc} 0 & 0 & 0 & 1 \\ 1 & 0 & 0 & 0 \\ 0 & 1 & 0 & 0 \\ 0 & 0 & 1 & 0 \end{array}\right), $$  commutes with the link operator $C_l$. This is seen to be true from the definition of the link operator. Consider
\bea X_l^2C_l|\chi_{v_1}, \phi_l, \chi_{v_2}\rangle & = & \langle\mu(\phi_l).\chi_{v_1}|\chi_{v_2}\rangle X_l^2|\chi_{v_1}, \phi_l, \chi_{v_2}\rangle \\ & = &  \langle\mu(\phi_l).\chi_{v_1}|\chi_{v_2}\rangle |\chi_{v_1}, \phi_{l+2}, \chi_{v_2}\rangle \eea whereas on the other hand we have 
\bea C_lX^2_l|\chi_{v_1}, \phi_l, \chi_{v_2}\rangle & = &  \langle\mu(\phi_{l+2}).\chi_{v_1}|\chi_{v_2}\rangle |\chi_{v_1}, \phi_{l+2}, \chi_{v_2}\rangle \\ & = &   \langle\mu(\phi_l).\chi_{v_1}|\chi_{v_2}\rangle |\chi_{v_1}, \phi_{l+2}, \chi_{v_2}\rangle \eea
as $\phi_2$ acts trivially on $H_2$. Thus in this model we have deconfined an extra flux by a mere choice of representation space for the gauge group.

\subsection{Models with Deconfined Gauge Excitations}

The link operator $C_l$ leads to confinement of the flux excitations in the previous two examples. Thus if we try to move a pair of fluxes apart we spend energy coming from the link excitations. The simplest way to avoid this is by making the link operator identity on all the links of the lattice, leaving only vertex and plaquette terms. The Hamiltonian for the $H_2/\mathbb{Z}_2$ case now becomes 
\beq H = -\sum_v A_v^1 - \sum_p B_p^1 \eeq
where $A_v^1$ is given by Eq.(\ref{av1}) and $B_p^1$ is given by Eq.(\ref{bp1}) respectively. This resembles the toric code Hamiltonian with the addition of the matter fields on the vertices. It is easy to see that the fluxes are no longer confined in this case as there is no cost in energy in moving a pair of fluxes apart. The charge excitations are not confined as before. We have one vertex excitation as before on a vertex $v$ which appears by acting with $\sigma^z_v$ on the ground state.

We can carry out this construction for other gauge groups as well after choosing appropriate vector spaces for $H_n$ which carry the representation of these gauge groups. If we turn off the effect of the link operator $C_l$ in all these cases we retain the deconfined flux excitations of the corresponding quantum double. 

\subsection{Recovering the Quantum Double Hamiltonians of Kitaev}

If we choose the vector space on which the gauge group acts to be the trivial one dimensional vector space $H_1$, the gauge group acts trivially on such a space. This makes all the operators act non-trivially only on the gauge sector and effectively ``switches off'' the matter sector as the operators act trivially on this sector by just scalar multiplication. In particular the link operator $C_l$ now becomes identity by construction. Thus we are left with only the vertex operator and the plaquette operators which act only on the gauge fields reproducing precisely the quantum double Hamiltonians of Kitaev. 
  
\section{Outlook}

We have presented a systematic method to construct the transfer matrices of two dimensional lattice theories with gauge and matter fields inspired by the state sum construction of Kuperberg~\cite{Kuper1}. The construction is a an extension of the one we started in~\cite{pp1} by including matter fields. From the mathematical point of view the construction in~\cite{pp1} produced quantum doubles of various inputs of which gauge group algebras are special cases. In this paper we have more than a quantum double of these inputs due to the inclusion of the matter fields. If we use weak Hopf algebras and their modules as inputs in the construction presented in this paper we can obtain the Levin-Wen string-net models with confined excitations which was seen to be a feature in these models due to the presence of the link operator $C_l$. 

The exactly solvable models obtained for certain parameters in this paper were all sums of commuting projectors. They described fully interacting lattice theories with gauge and matter fields in two dimensions for arbitrary involutory Hopf algebras. We showed examples with long-ranged entangled ground states in this paper. We also expect to find models with short-ranged entangled states in this parameter space or by cleverly enlarging it to obtain the honeycomb model of Kitaev~\cite{KitHoney}. 

Moving from the ground state sector we find interesting excitations in these models. They contain both vertex and gauge excitations. While the vertex excitations are not anyons in the usual sense they nevertheless have interesting properties as was studied in~\cite{pp2} where we showed that one can obtain non-Abelian fusion rules from a system with just Abelian degrees of freedom. While the charge excitations are deconfined as in the quantum double case, we have confined flux excitations in these models. The confinement arises due to the introduction of link operators which are symmetric under the local and global symmetries. Thus we have exactly solvable models with confined and deconfined particles with the latter protected by the energy gap created by these link operators. We will elaborate more on these models in a forthcoming publication.

\section*{Acknowledgements}

PTS, PP and MJBF would like to thank FAPESP for support of this work. JPIJ thanks CNPq for the financial support. PP also thanks Sanatan Digal for hospitality in IMSc, Chennai where part of this work was done. PP also thanks S. R. Hassan and G. Baskaran for useful discussions during his stay in IMSc, Chennai.   
\pagebreak
\appendix
\section{Input Data for the Construction}
The algebraic structures needed to construct the models comprise an involutory Hopf Algebra $\langle\mathcal{A},m,\eta, \Delta, \epsilon, S  \rangle$ and a left $n$-dimensional $\mathcal{A}$-module $\langle H, \mu \rangle$ over a field $\mathbb{C}$ that is equipped with a co-structure $t$ and a bilinear form $G$. We briefly describe each algebraic structure together with the properties relevant for the construction of the models. The notation we use is similar to the one used in~\cite{pp1} which was first introduced by \emph{Kuperberg} in~\cite{Kuper1}.

\subsection{Hopf Algebras}
Let $\mathbb{C}$ be a field and $\mathcal{A}$ be an $n$-dimensional vector space over this field. Denote the basis by $\{\phi_i\}_{i=1}^{n}$ and its dual basis by $\{\phi^i\}_{i=1}^n$, defined such that $\phi^i(\phi_j)=\delta_j^i $.

\begin{figure}[h!]
\centering
\includegraphics[scale=1]{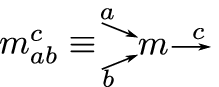}
\caption{Kuperberg diagram for the multiplication map $m$.} \label{fig:mhopf}
\end{figure}

The vector field $\mathcal{A}$ is said to be an algebra if there are two linear maps:
$$m:\mathcal{A}\otimes \mathcal{A} \rightarrow \mathcal{A}\qquad \text{and}\qquad \eta:\mathbb{C}\rightarrow \mathcal{A},$$
where $m$ is an associative multiplication map and $\eta(1)$ is the unit element. The multiplication map is defined by its action on the basis 
elements as follows,
\begin{equation}\label{mhopf}
 m(\phi_a\otimes\phi_b)= m_{ab}^c\phi_c,
\end{equation}
where the sum over repeated indices is implied. The coefficients $m_{ab}^c$ are called \emph{structure constants} and can be thought of as tensors representing elements of $\mathcal{A}\otimes\mathcal{A}\otimes\mathcal{A}^{\ast}$ so it is natural to associate them to the Kuperberg diagram on figure (\ref{fig:mhopf}).

\begin{figure}[h!]
 \centering
 \includegraphics[scale=1]{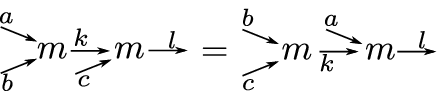}3
 \caption{Associative property of the multiplication map.}\label{fig:associa}
\end{figure}

\begin{figure}[h!]
 \centering
 \includegraphics[scale=1]{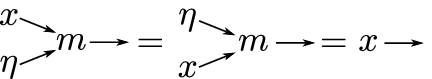}
 \caption{Existence of the unit element for the multiplication map.}\label{fig:unit}
\end{figure}

Also, we require the multiplication map to be associative such that the product of three basis elements $\phi_a$, $\phi_b$ and $\phi_c$ in $\mathcal{A}$ is: 
\begin{displaymath}
 (\phi_a\phi_b)\phi_c=\phi_a(\phi_b\phi_c),
\end{displaymath}
which can be expressed in terms of the structure constants as follows
\begin{equation}
 m_{ab}^{k}m_{kc}^{l}=m_{ak}^{l}m_{bc}^{k},
\end{equation}
Alternatively we can represent this associative property of the multiplication map using a Kuperberg diagram as shown in figure (\ref{fig:associa}).
\\
The unit element $\eta\in \mathcal{A}$ is such that for all $x \in \mathcal{A}$, $x\eta=\eta x=x$ as depicted in figure (\ref{fig:unit}). 
\\
If both conditions shown in figures (\ref{fig:associa}) and (\ref{fig:unit}) are fulfilled we say the triple $\langle \mathcal{A},m,\eta\rangle$ 
forms an associative algebra with unit element $\eta$.\\
Similarly, we can take the dual vector space $\mathcal{A}^{\ast}$ and define a multiplication map on it, $\Delta:\mathcal{A}^{\ast} \otimes \mathcal{A}^{\ast} \rightarrow \mathcal{A}^{\ast}$ by means of the action on two elements of the dual basis $\{\phi^i\}_{i=1}^n$, 
\begin{equation}
 \Delta(\phi^i \otimes \phi^j)=\phi^i \phi^j = \Delta^{ij}_{k} \phi^k ,
\end{equation}
where the coefficients $\Delta^{ij}_{k}$ are the \emph{structure constants}. There is an unit element that can be seen as a map
$\epsilon: \mathbb{C} \rightarrow \mathcal{A}^{\ast}$ such that $\epsilon(1)$ is the unity for the dual multiplicative map $\Delta$.
So the triple $\langle\mathcal{A}^{\ast},\Delta,\epsilon\rangle$ defines an algebra structure in $\mathcal{A}^{\ast}$; Equivalently we can regard 
the maps $\Delta$ and $\epsilon$ as being a co-multiplication and co-unity in $\mathcal{A}$ respectively. In this sense, the map 
$\Delta:\mathcal{A}\rightarrow \mathcal{A}\otimes\mathcal{A} $ is defined as
\begin{equation}
 \Delta(\phi_i)=\Delta_i^{jk}(\phi_j \otimes \phi_k),
\end{equation}
and the unit map $\epsilon: \mathcal{A}\rightarrow\mathbb{C}$ is defined as
\begin{equation}
 \epsilon(\phi_i)=\epsilon^i.
\end{equation}
The triple $\langle\mathcal{A},\Delta,\epsilon \rangle$ forms a co-associative co-algebra with co-unit provided the following two relations are 
fulfilled
\begin{figure}[h!]
 \centering
 \subfigure[ ]{\includegraphics[scale=1]{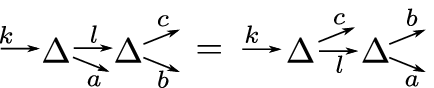}}\qquad \qquad 
 \subfigure[ ]{\includegraphics[scale=1]{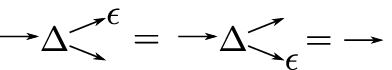}}
 \label{fig:coalg}\caption{Associativity of the co-multiplication map and existence of the co-unit of $\Delta$.}
\end{figure}

\begin{figure}[h!]
\centering
\subfigure[]{\includegraphics[scale=1]{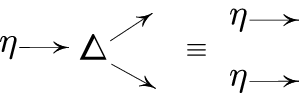}}\qquad \subfigure[]{\includegraphics[scale=1]{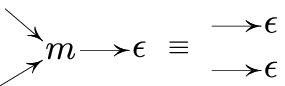}}\\
\subfigure[]{\includegraphics[scale=1]{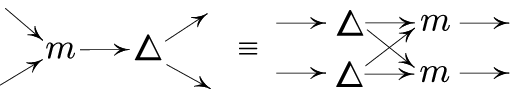}}
\caption{Bi-algebra compatibility conditions.}\label{fig:bialgebra}
\end{figure}

The quintet $\langle\mathcal{A},m,\eta,\Delta,\epsilon\rangle$ is said to form a bi-algebra whenever some special compatibility conditions are satisfied. This is, when the co-multiplication and the co-unit map are homomorphisms of the algebra, namely
\begin{align}
\Delta(\phi_a\phi_b)&=\Delta(\phi_a)\Delta(\phi_b), \label{bialg1}\\
\epsilon(\phi_a\phi_b)&=\epsilon(\phi_a)\epsilon(\phi_b).\label{bialg2}
\end{align}
The Kuperberg diagrams for these conditions are shown in figure (\ref{fig:bialgebra}).

\begin{figure}[h!]
\centering
\includegraphics[scale=1]{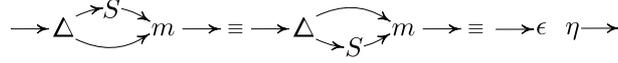}
\caption{The antipode condition.}\label{fig:antipode}
\end{figure}

Consider now the endomorphism $S:\mathcal{A}\rightarrow\mathcal{A}$ called the antipode of the algebra. If such a map satisfies the condition shown in figure (\ref{fig:antipode}) then we say the sextet $\langle\mathcal{A},m,\eta, \Delta, \epsilon, S  \rangle$ forms a Hopf Algebra.  
Additionally the sextet can be equipped with a $\ast$-structure which ensures the definition of Hilbert spaces over complex numbers and unitarity of the system \cite{kassel,Agu}. Consider the conjugate linear involution $\ast:\mathcal{A}\rightarrow \mathcal{A}$ satisfying
\begin{align}
(x^\ast)^\ast&=x,\\
(xy)^\ast&=y^\ast x^\ast,\\
\eta^\ast&=\eta
\end{align}
for all $x,y \in \mathcal{A}$ and the unit element $\eta \in \mathcal{A}$. Then the triplet $\langle\mathcal{A},m, \eta \rangle$ is called a $\ast$-algebra. Moreover, the quintet $\langle\mathcal{A},m,\Delta,\eta,\epsilon \rangle$ is called a $\ast$-bialgebra if the involution map $\ast$ is consistent with the co-product structure, i.e.
\begin{align}
\Delta(x^\ast)&=\Delta(x)^\ast,\\
\epsilon(x^\ast)&=\overline{\epsilon(x)}, 
\end{align}
where the bar in the r.h.s. of the last expression stands for the complex conjugate. Finally, the sextet $\langle\mathcal{A},m,\Delta,\eta,\epsilon,S \rangle$ is called a Hopf $\ast$-algebra if the conjugate linear involution $\ast$ is compatible with the antipode $S$ in the following sense:
\begin{equation}
S(S(x)^\ast)^\ast=x.
\end{equation}

\subsection{Left $\mathcal{A}$-Module}
Consider the triple $\langle\mathcal{A},m,\eta \rangle$ as being an algebra over the field $\mathbb{C}$, let $H_n$ also be a vector space over the same field $\mathbb{C}$, and let $\mu:\mathcal{A}\otimes H_n \rightarrow H_n$ be a morphism of vector spaces such that it takes an element $g\in\mathcal{A}$ and acts with it on an element $\chi\in H_n$, we denote this action as $\mu(g)\triangleright\chi$. The pair $\langle H_n,\mu\rangle$ is said to be a \emph{left $\mathcal{A}-$module}~\cite{anderson, james} if the following properties are satisfied for all $\chi,\chi^{\prime} \in H_n $; $c \in \mathbb{K}$, $g,h \in \mathcal{A} $ and $\eta$ the unit element of the algebra.

\begin{align}
 \mu(g)\triangleright \chi &\in H_n, \label{P1}\\
 \mu(hg) \triangleright  \chi &= \mu(h)\triangleright \left( \mu(g) \triangleright \chi\right), \label{P2}\\
 \mu(\eta)\triangleright \chi  &= \chi, \label{P3}\\
 \mu(g)\triangleright (c \chi)  &= c \left( \mu(g) \triangleright \chi   \right) \label{P4}\\
 \mu(g) \triangleright (\chi + \chi^{\prime} )  &= \mu(g)  \triangleright \chi   + \mu(g) \triangleright \chi^{\prime} \label{P5}
\end{align}

\begin{figure}[h!]
\centering
  \includegraphics[scale=1]{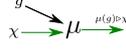}
  \caption{Diagram associated to the module structure.}\label{fig:module1}
\end{figure}

Let us associate the diagram in figure (\ref{fig:module1}) to the new structure constant of the \emph{left $\mathbb{C}\mathcal{A}-$Module}. Note that we make a distinction between the kind of arrows in figure (\ref{fig:module1}). This is due to the fact that the module map $\mu$ combines elements of two different spaces. Consequently  we use a green arrow to represent an element of the vector space $H_n$ while a black arrow represents an element of the vector space $\mathcal{A}$. 

\begin{figure}[h!]
\centering
\subfigure[Eq.~\ref{P2}.]{\includegraphics[scale=1]{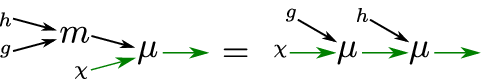}}\qquad
\subfigure[Eq.~\ref{P3}.]{\includegraphics[scale=1]{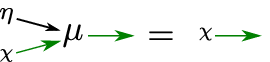}}\\
\subfigure[Eq.~\ref{P4}.]{\includegraphics[scale=1]{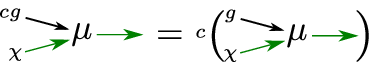}}\qquad
\subfigure[Eq.~\ref{P5}.]{\includegraphics[scale=1]{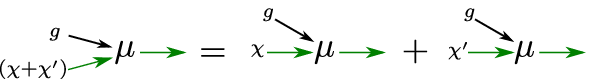}}
\caption{Diagrammatic notation of properties \ref{P2} to \ref{P5}.}\label{fig:modprop} 
\end{figure}

Using this diagrammatic representation we can re-write properties in Eqs.~(\ref{P1}) to (\ref{P5}) as depicted in figure (\ref{fig:modprop}).

\begin{figure}[h!]
\centering
\subfigure[]{\includegraphics[scale=1]{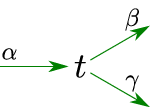}} \qquad \qquad \subfigure[]{\includegraphics[scale=1]{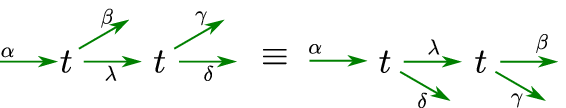}}
\caption{(a) The co-structure tensor. (b) Associativity of the co-product}\label{fig:t}
\end{figure}

We equip the pair $\langle H_n,\mu \rangle$ with a co-associative co-multiplication structure given by the map $t: H_n\rightarrow H_n\otimes H_n$. If $\{\chi_{\alpha}\}_{\alpha=1}^n$ are the basis elements of $H_n$ the co-multiplication map is defined by the action on these elements as follows:
$$t(\chi_{\alpha})= t^{\beta\gamma}_{\alpha} (\chi_\beta \otimes \chi_\gamma),$$ 
where $t^{\beta\gamma}_{\alpha}$ are the structure constants and are represented by the Kuperberg diagram in figure (\ref{fig:t}(a)). Moreover, the co-structure is co-associative meaning that the condition depicted in figure (\ref{fig:t}(b)) is satisfied.

\begin{figure}
\centering
\includegraphics[scale=1]{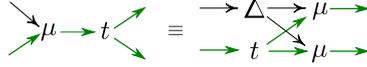}
\caption{The compatibility relation between the left $\mathcal{A}$-module and the co-product of the Hopf algebra.}\label{fig:hopfmod}
\end{figure}

There is a compatibility relation between the left $\mathcal{A}$-module and the co-product of the Hopf algebra $\mathcal{A}$ shown in figure (\ref{fig:hopfmod}). It implies that the co-structure map $t$ preserves the action of the $\mathcal{A}$-module, i.e.,
\begin{equation}\label{hopfmod}
t(\mu(a)\triangleright \chi)\equiv \mu(\Delta(a))\triangleright t(\chi)
\end{equation}

\begin{figure}[h!]
\centering
\includegraphics[scale=1]{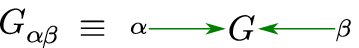}
\caption{The bilinear map G and its tensor representation.}\label{fig:Gmap}
\end{figure}

Finally, we define a bilinear map $G: H_n\otimes H_n \rightarrow \mathbb{K}$. It can be represented by a tensor with two incoming green arrows as shown in figure (\ref{fig:Gmap}). This tensor will help in contracting the arrows for the matter degrees of freedom when constructing the partition function of the model.

The above algebraic structures are the elementary building blocks of the partition function and the transfer matrix from which we will obtain specific models written in terms of local operators.

\section{Splitting of the (1+1)D Transfer Matrix}
\subsection{Associating Tensors to the Lattice}\label{sec:tensors}

\begin{figure}[h!]
\centering
\includegraphics[scale=1]{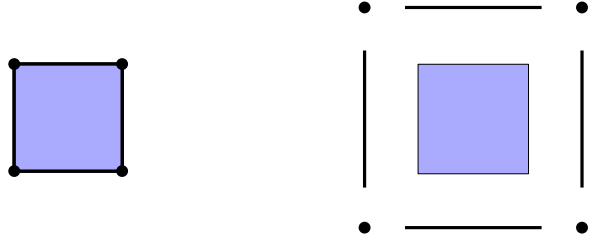}
\caption{A $(1+1)D$ square lattice cell and its components, vertices, links and face are shown.}\label{fig:lattcomp}
\end{figure}

In this case we consider a 2-dimensional lattice $\mathcal{L}$ composed of vertices, links and faces as shown in figure (\ref{fig:lattcomp})

As in \S~(1.1.1) to each of the lattice components we associate a set of tensors. The $M_{a_1a_2a_3a_4}$ tensor is associated to the faces of the lattice. Since each vertex $\mathcal{L}$ is now four-valent a 4-tensor $T^{\alpha_1 \alpha_2 \alpha_3 \alpha_4}$ is associated to it. To the links of the lattice we associate a $L^{ab\beta}_{\alpha}$ tensor, as each link is connected to two vertices and two faces. This is schematized in figure (\ref{fig:tensors1}).
\begin{figure}[h!]
\centering
\subfigure[]{
\includegraphics[scale=1]{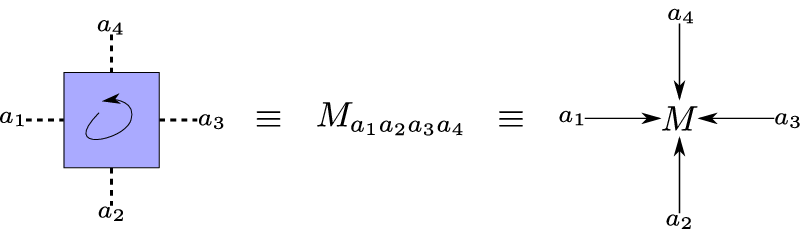}} \\
\subfigure[]{\includegraphics[scale=1]{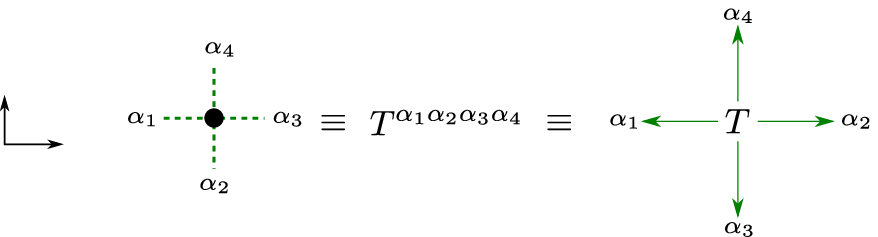}}\label{fig:tensors1}\\
\subfigure[]{\includegraphics[scale=1]{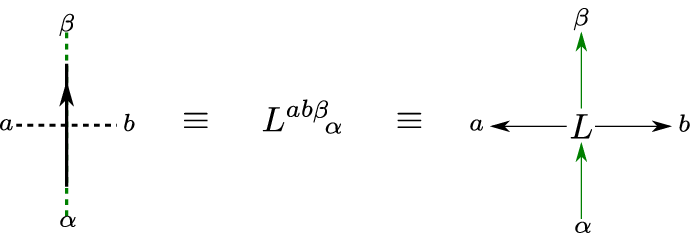}}
\caption{The tensors associated to each component of the lattice.}\label{fig:tensors1}
\end{figure}

The $M_{a_1a_2a_3a_4}$ tensor involves gauge degrees of freedom only and consequently has Latin indices only. The $T^{\alpha_1 \alpha_2 \alpha_3 \alpha_4}$ involves only matter degrees of freedom therefore it has only Greek indices. Consequently, the mixed tensor $L^{ab\beta}_{\alpha}$ involves contractions between matter and gauge degrees of freedom. The contraction of these tensors is orientation dependent. The antipode of the Hopf algebra $S:\mathcal{A}\rightarrow\mathcal{A}$ takes care of the orientation in the gauge sector whereas the bilinear form $G:H\otimes H \rightarrow {C}$ performs the task for the matter sector. Thus, the contraction rules are the same as in the $(2+1)D$ case. For the gauge degrees of freedom an antipode is placed whenever the orientation of the plaquette does not match the one of the link. Similarly for the matter degrees of freedom, the green arrow coming out from the mixed tensor $L^{ab\beta}_{\alpha}$ is joined to the $T^{\alpha_1 \alpha_2 \alpha_3 \alpha_4}$ via the bilinear form $G_{\beta\alpha_1}$.

\subsection{Splitting the Transfer Matrix}
\label{sec:splitting}
The splitting procedure shown in \S~\ref{sec:split} can, as well, be used to write the transfer matrix of the $(1+1)D$ case as a product of local operators. 
The result of contracting all the tensors on the lattice is defined to be the partition function of the theory, given by:
\begin{equation}
Z(\mathcal{L},\mathcal{A},H,z,\xi)=\prod_{p}M_{a_1a_2a_3a_4}(p)\prod_{v}T^{\alpha_1\alpha_2\alpha_3\alpha_4}(v)\prod_{l}L^{ab\beta}_{\alpha}(l)\prod_{o}S^a_b\prod_{l}G_{\alpha \beta}.
\end{equation}
\begin{figure}[h!]
\centering
\includegraphics[scale=1]{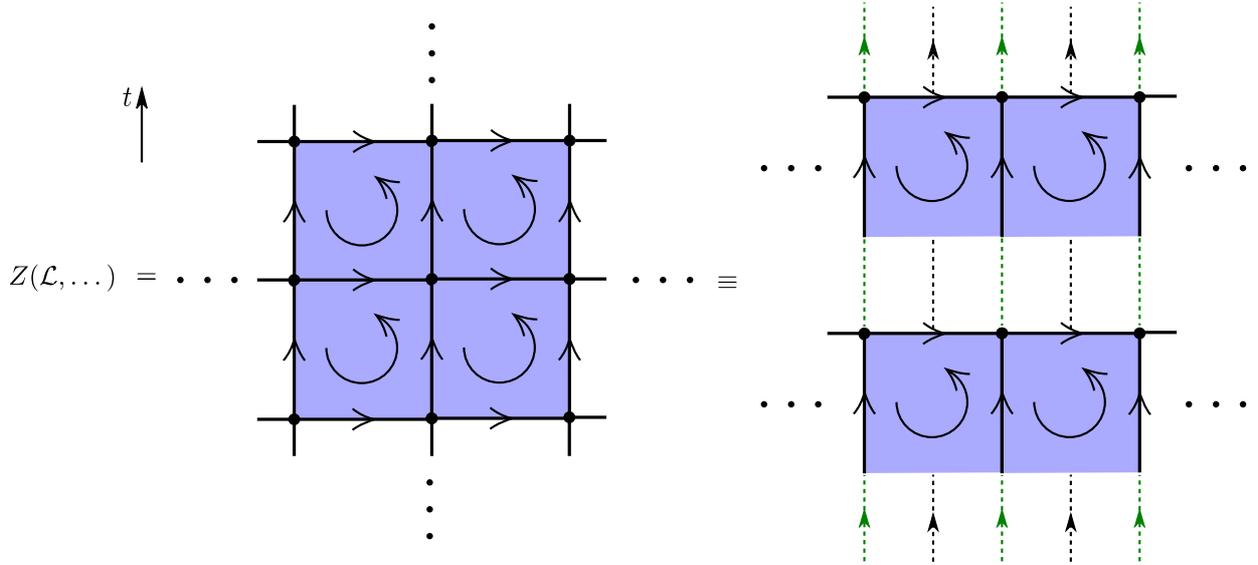}
\caption{The graphical representation of the partition function $Z(\mathcal{L},\mathcal{A},H,z,\xi)$ is shown.}
\label{fig:Z}
\end{figure}

The partition function is related to the one step evolution operator $U$ as follows:
\begin{equation}
Z=tr(U^N),
\end{equation}
therefore $U$ is represented as a single slice on the graphical representation of figure (\ref{fig:Z}).
\begin{figure}[h!]
\centering
\includegraphics[scale=1]{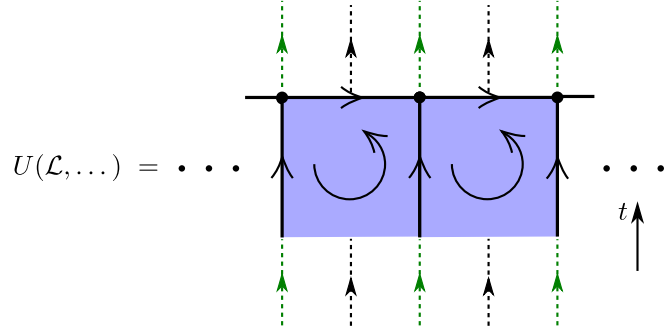}
\caption{The one step evolution operator $U(\mathcal{L},\mathcal{A},)$ is one slice of the partition function.}
\label{fig:U1}
\end{figure}

From the correspondence shown in figure (\ref{fig:tensors1}) the one step evolution operator can be written as a tensor network built from the tensors representing the structure constants of the algebra $\mathcal{A}$, the module $H$, the antipode $S$ and the bilinear form $G$. This results in a rather intricate representation that will be split in operators acting on smaller pieces of the total Hilbert space, as we will we show later on.

\begin{figure}[h!]
\centering
\includegraphics[scale=1]{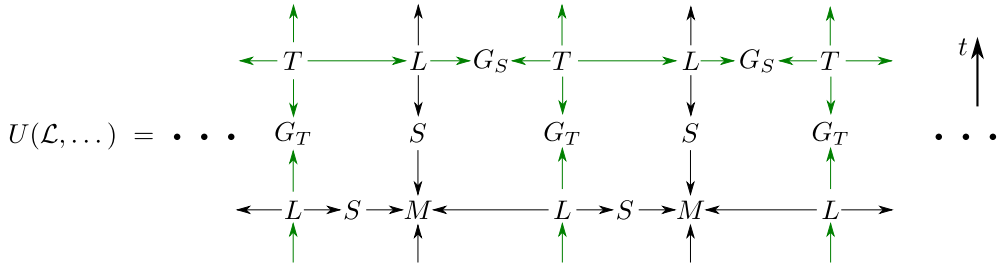}
\caption{The tensor network representation of the one step evolution operator is shown. }\label{fig:UTN}
\end{figure}

\begin{figure}[h!]
\centering
\includegraphics[scale=1]{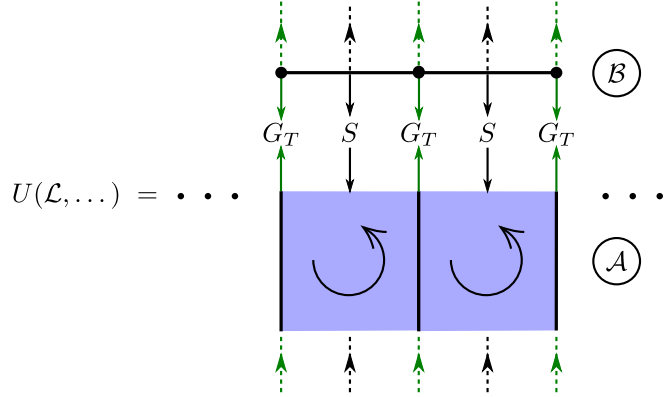}
\caption{Step 1 of the splitting procedure}\label{fig:step1}
\end{figure}

\begin{figure}[h!]
\centering
\includegraphics[scale=1]{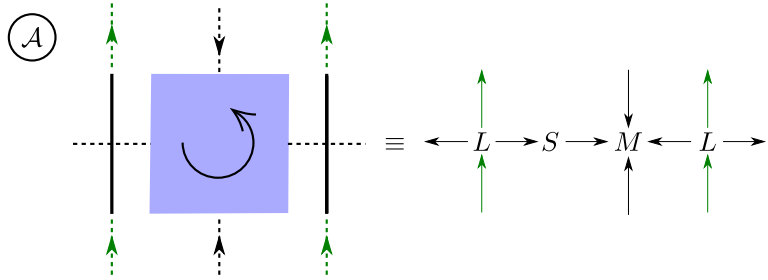}
\caption{Tensor network representation of a plaquette and the two adjacent timelike links.}\label{fig:step2}
\end{figure}

The process of writing the one step evolution operator $U$ as a product of local operators is based on the algebraic properties of the tensors that represent the elements of the lattice. We begin by splitting the diagram in figure (\ref{fig:U1}) into two parts which we call $\mathcal{A}$ and $\mathcal{B}$ as shown in figure (\ref{fig:step1}). 
Now we proceed to the splitting of the $\mathcal{A}$ part. It is sufficient to pick one of the plaquettes and the two adjacent timelike links and write the tensor network representation. This is depicted in figure (\ref{fig:step2}). 

\begin{figure}[h!]
\centering
\subfigure[]{\includegraphics[scale=1]{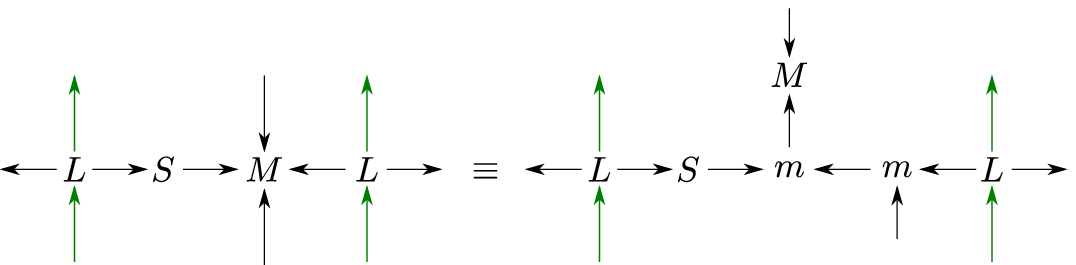}} \\
\subfigure[]{\includegraphics[scale=1.2]{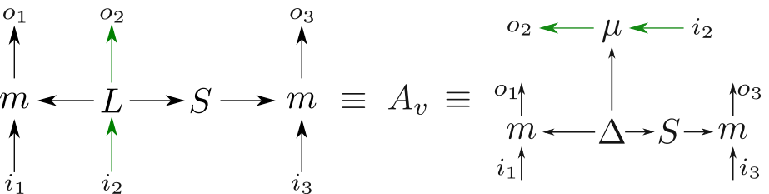}}
\caption{(a) Splitting of the $M$ tensor. (b) The vertex operator $A_v$.}\label{fig:step3}
\end{figure}

The associative property of the multiplication map $m$ allows us to literally split the $M$ tensor representing the plaquette of the lattice as in figure (\ref{fig:step3}(a)). At this point, we are able to define the vertex operator as the tensor network shown in figure (\ref{fig:step3}(b)) such that it acts on two links denoted $i_1$, $i_3$ and the vertex $i_2$ in between. 

\begin{figure}[h!]
\centering
\includegraphics[scale=1]{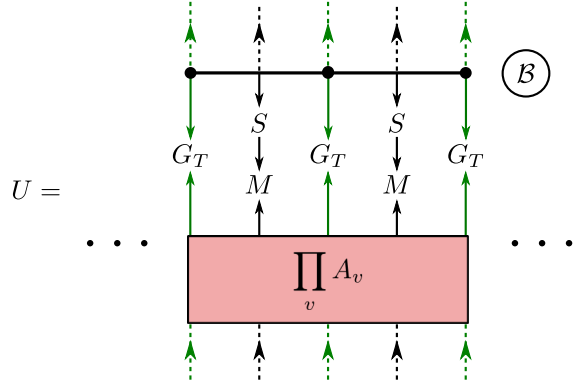}
\caption{The $U$ operator after the splitting of the $\mathcal{A}$ part of the tensor network in figure (\ref{fig:step1}).}\label{fig:U3}
\end{figure}

Therefore, the $U$ operator now can be represented as a product of vertex operators $A_v$ for all vertices $v$ in the lattice, as depicted in figure (\ref{fig:U3})

\begin{figure}[h!]
\centering
\subfigure[]{\includegraphics[scale=1]{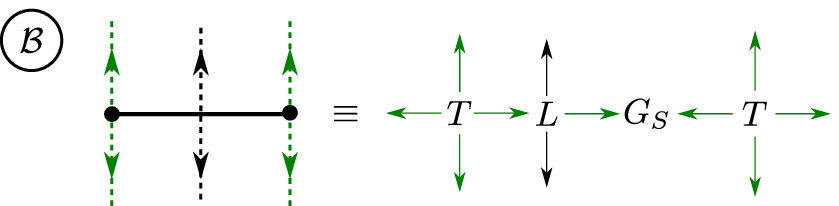}}
\subfigure[]{\includegraphics[scale=1]{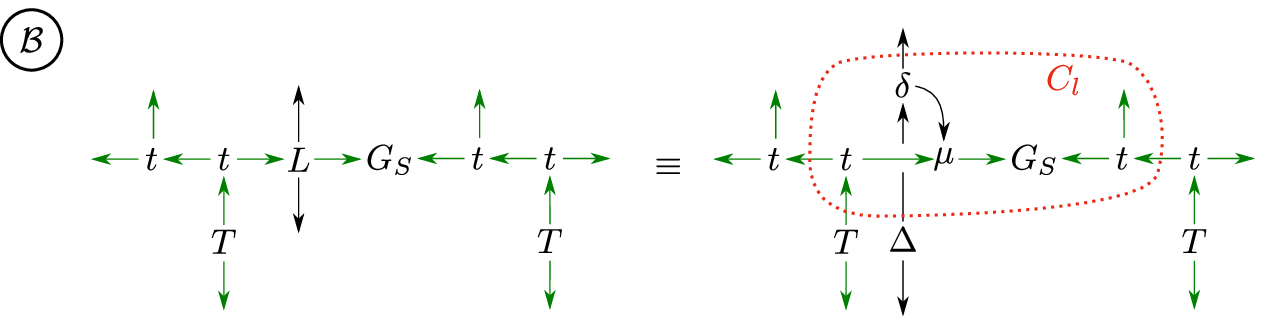}}
\caption{In (a) we show the tensor network representation of a pair of vertices and the spacelike link in between. In (b) the link operator is explicitely shown.}\label{fig:step5}
\end{figure}

Similarly, we now proceed to the splitting of the upper section in figure (\ref{fig:step1}) that we called $\mathcal{B}$. It is enough to consider two vertices and the spacelike link in between, consequently the TN representation is shown in figure (\ref{fig:step5}).

\begin{figure}[h!]
\centering
\includegraphics[scale=1]{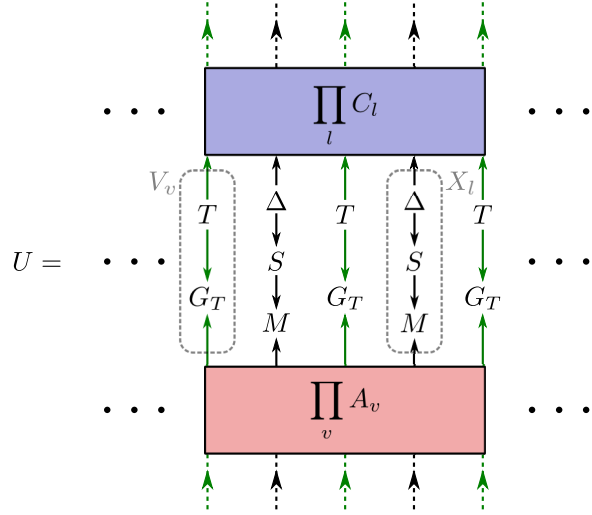}
\caption{Final form of the $U$ operator after the systematic split of the TN performed over the entire lattice.}\label{fig:step6}
\end{figure}

The associative nature of the co-multiplicative map $t$ allows us to split the $T$ tensors on each vertex as shown in the l.h.s of figure (\ref{fig:step5}(b)) by writing the mixed operator $L^{ab \beta}_{\alpha}$ in terms of its elementary components we get the link operator $C_l$ as the TN highlighted in the r.h.s. of figure (\ref{fig:step5}(b)). Thus, the $\mathcal{B}$ part of the $U$ operator is now written as a product of $C_l$ for each link of the lattice together with the connector operators which we call $L_l$ and $V_v$ that act on single spacelike links and vertices of the lattice, respectively. Finally the one step evolution operator has been decomposed as a product of the local operators depicted in figure (\ref{fig:step6}).

\begin{figure}[h!]
\centering
\subfigure[]{\includegraphics[scale=1.2]{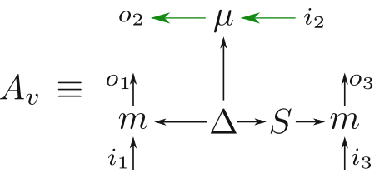}} \\
\subfigure[]{\includegraphics[scale=1.2]{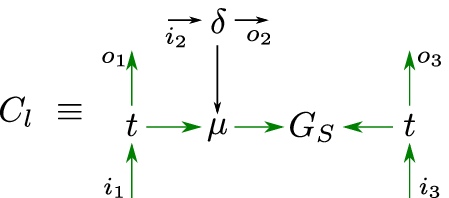}}\\
\subfigure[]{\includegraphics[scale=1.2]{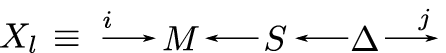}}\\
\subfigure[]{\includegraphics[scale=1.2]{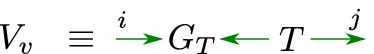}}
\caption{The tensorial representations of each local operator that make up the one step evolution operator $U$.}\label{fig:ops}
\end{figure}

\begin{figure}[h!]
\centering
\subfigure[operator which acts on a single vertex.]{
\includegraphics[scale=1]{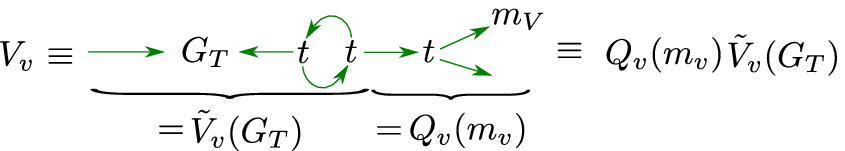} \label{linkvertice-a}
}
\hspace{.5cm}
\subfigure[operator which acts on a single link.]{
\includegraphics[scale=1]{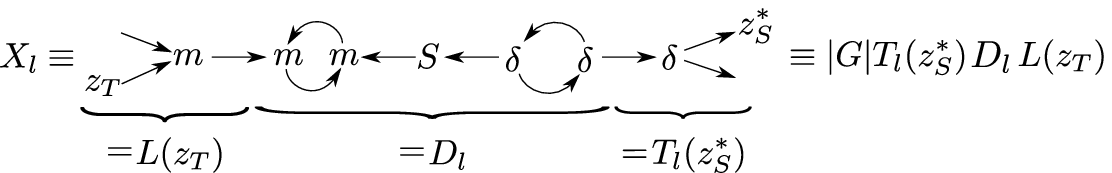} \label{linkvertice-b}
}
\caption{Operators which act on the links and vertex of the lattice.}
\label{linkvertice}
\end{figure}

Therefore, the complete transfer matrix of the theory can be written as:
\begin{equation}
U\left(z_\textrm{T},z_\textrm{S}^*,z_\textrm{T}^*,m_V,G_S,G_T \right)=\prod_l C_l(G_S) \prod_l X_l \prod_v V_v \prod_v A_v(z_T^*)
\end{equation}
where the tensor network representation of each local operator is shown in figure (\ref{fig:ops}).
The operators $V_v$ and $L_l$ can still be written as shown in figure (\ref{linkvertice}), in which we have used the operators defined in figure (\ref{LRTQops}) and the antipode identity for involutory Hopf algebra.
Finally we can write the transfer matrix as
$$U(z_T,z_S^*,z_T^*,m_v,G_S)=\prod_l C_l(G_S) \prod_v Q_v(m_v) \prod_l L_l(z_T)T_l(z_S^*) \prod_v A_v(z_T^*)\;.$$
The operator $D_l$ can shown to be proportional to identity, more precisely $D_l =\vert G \vert \mathbb{I}$, while the operator $\tilde{V}_v(G_T)$ can be made into identity by choose $(G_T)_{\alpha\beta}=\delta(\alpha,\beta)$.

\section{Examples of Exactly solvable $1$-D Quantum Models}

In these examples there are no plaquette operators as there are no plaquettes in one dimension. The models we consider will be made of only the vertex operator $A_v$, which are again the gauge transformations, and the link operators $C_l$ which describe the gauge and the matter interactions. Thus our Hamiltonians will be of the form 
\beq H = -\sum_v A_v -\sum_lC_l .\eeq

We will just look at two examples, $H_2/\mathbb{Z}_2$ and $H_2/\mathbb{Z}_4$. We will see that in the first case there are deconfined vertex and flux excitations with no ground state degeneracy. In the second case there continues to be deconfined vertex excitations but the ground state in this case has a lot of degeneracy which we attribute to the ``condensation'' of certain fluxes of the corresponding model in two dimensions. 

\subsection{$H_2/\mathbb{Z}_2$:}  

The action of $\mathbb{Z}_2$ on $H_2$ is the same as in the two dimensional example. 

The vertex operator is given by 
\beq A_v = \frac{1 + \sigma^x_{l_1}\sigma_v^x\sigma^x_{l_2}}{2}\eeq
where $l_1$ and $l_2$ are the two links adjacent to the vertex $v$. 

The link operator is given by 
\beq C_l = \frac{1 + \sigma^z_{v_1}\sigma_l^z\sigma^z_{v_2}}{2}\eeq
where $v_1$ and $v_2$ are the two vertices flanking the link $l$. 

It is easy to see that these operators commute with each other and are projectors. Thus as before the Hamiltonian is exactly solvable and its spectrum is easily obtained. the ground state conditions are similar to the corresponding example in the two dimensional case which is $A_v|gr\rangle = C_l|gr\rangle=|gr\rangle,~\forall v,l$. Using the arguments in the two dimensional case the number of ground states is again one for this model. They are constructed in manner similar to the way it was done for the two dimensional case. They are given by
\bea |\psi_v\rangle & = & \prod_v A_v\otimes_l|\phi_1\rangle\otimes_v|\chi_1\rangle \\ 
|\psi_l\rangle & = & \prod_l C_l \otimes_l|\phi_1+\phi_{-1}\rangle\otimes_v|\chi_1+\chi_{-1}\rangle. \eea
As before we expect these two states to be the same. 

\begin{figure}[h!]
\centering
\includegraphics[scale=1]{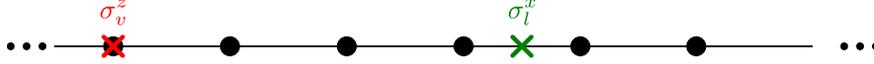}
\caption{The isolated vertex and link excitations in the $H_2/\mathbb{Z}_2$ model.}
\label{vertex1p1}
\end{figure}

\begin{figure}[h!]
\centering
\includegraphics[scale=1]{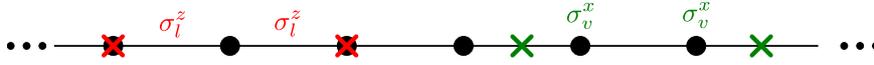}
\caption{The deconfined vertex and link excitations in the $H_2/\mathbb{Z}_2$ model.}
\label{gauge1p1}
\end{figure}

The excitations are obtained as before by applying $\sigma^x$ to either the vertices or the links to create link excitations. The former gives us deconfined link excitations and the latter gives us isolated link excitations. By applying $\sigma^z$ to either vertices of links we obtain the vertex excitations. The former gives us isolated vertex excitation whereas the latter gives us deconfined charge excitations. The isolated vertex and link excitations are shown in figure (\ref{vertex1p1}) and the deconfined gauge excitations are shown in figure (\ref{gauge1p1}) respectively.

\subsection{$H_2/\mathbb{Z}_4$:}

The gauge action of $\mathbb{Z}_4$ on $H_2$ is taken to be the same as the one in the two dimensional case. The novelty in one dimension is an interesting phenomena by which the ground state degeneracy increases. This is because a single operator $X_l^2$ on a link $l$ commutes with the Hamiltionian. This was shown in the corresponding example in the two dimensional case, the difference here is that there is no plaquette operator in one dimension. Note that this is not a gauge transformation as then we would have to act on the adjacent vertex and link of $l$ as well. Thus we have effectively ``condensed'' the two dimensional fluxes obtained by applying the operator $X_l^2$ along a path in the dual lattice. The deconfined charge excitations still exist thus giving us a model with deconfined excitations and ground state degeneracy more than one in one dimensions.

\end{document}